\documentclass[journal]{IEEEtran}

\usepackage{array}
\usepackage{enumerate}
\usepackage{graphicx}
\usepackage{svg}
\usepackage{amsmath}
\usepackage{dingbat}
\usepackage{booktabs}
\usepackage{subcaption}
\usepackage{hyperref}
\usepackage{xcolor,soul}
\usepackage{makecell}
\usepackage{svg}
\newcommand{\xmark}{\ding{55}}%
\usepackage{pifont}
\usepackage{xcolor}

\DeclareUnicodeCharacter{2212}{-}

\begin{document}
\title{A Survey on Scalable LoRaWAN for Massive IoT: Recent Advances, Potentials, and Challenges}

\author{\IEEEauthorblockN{Mohammed Jouhari\IEEEauthorrefmark{1},
Nasir Saeed\IEEEauthorrefmark{2}, Mohamed-Slim Alouini\IEEEauthorrefmark{3}, and
El Mehdi Amhoud\IEEEauthorrefmark{1}
\\
}
\IEEEauthorblockA{\IEEEauthorrefmark{1}School of Computer Science, Mohammed VI Polytechnic University, Ben Guerir, Morocco\\
\IEEEauthorrefmark{2}Department of Electrical and Communication Engineering, United Arab Emirates University, Al Ain, UAE\\
\IEEEauthorrefmark{3}Computer, Electrical and Mathematical Sciences and Engineering Division (CEMSE), King Abdullah University of Science and Technology (KAUST), Thuwal
23955-6900, Saudi Arabia
\\
} 
 \{mohammed.jouhari, elmehdi.amhoud\}@um6p.ma, mr.nasir.saeed@ieee.org, slim.alouini@kaust.edu.sa}

\markboth{IEEE Communications Surveys \& Tutorials}%
{Shell \MakeLowercase{\textit{M. JOUHARI et al.}}: Bare Demo of IEEEtran.cls for IEEE Journals}

\maketitle

\begin{abstract}
Long-range (LoRa) technology is  most widely used for enabling low-power wide area networks (WANs) on unlicensed frequency bands. Despite its modest data rates, it provides extensive coverage for low-power devices, making it an ideal communication system for many internet of things (IoT) applications. In general, LoRa is considered as the physical layer, whereas LoRaWAN is the medium access control (MAC) layer of the LoRa stack that adopts a star topology to enable communication between multiple end devices (EDs) and the network gateway. The chirp spread spectrum modulation deals with LoRa signal interference and ensures long-range communication. At the same time, the adaptive data rate mechanism allows EDs to dynamically alter some LoRa features, such as the spreading factor (SF), code rate, and carrier frequency to address the time variance of communication conditions in dense networks. Despite the high LoRa connectivity demand, LoRa signals interference and concurrent transmission collisions are major limitations. Therefore, to enhance LoRaWAN capacity, the LoRa Alliance released many LoRaWAN versions, and the research community has provided numerous solutions to develop scalable LoRaWAN technology. Hence, we thoroughly examine LoRaWAN scalability challenges and state-of-the-art solutions in both the physical and MAC layers. These solutions primarily rely on SF, logical, and frequency channel assignment, whereas others propose new network topologies or implement signal processing schemes to cancel the interference and allow LoRaWAN to connect more EDs efficiently. A summary of the existing solutions in the literature is provided at the end of the paper, describing the advantages and disadvantages of each solution and suggesting possible enhancements as future research directions. 
\end{abstract}


\IEEEpeerreviewmaketitle

\section{Introduction}


The internet of things (IoT) is a revolutionary technology. It performs intelligent sensing and actuation for various objects by exchanging information with a core network.
People can manage or monitor the behavior of devices remotely from systems hundreds of kilometers away using various types of IoT technology. In academics and industry \cite{benaddi2022robust, benaddi2022anomaly}, IoT-based systems have proliferated in the last few years, providing multiple new applications such as smart homes, intelligent transportation, smart hospitals, and smart cities \cite{PueyoCentelles2021, Simo2021, Hoang2020}. Based on these emerging applications, the number of IoT devices is expected to grow from 7 billion in 2018 to 22 billion in 2025 \cite{Zhang2020}. Thus, massive IoT has emerged as a new paradigm, a category of IoT networks driven by scale, not communication speed. The number of connected devices in massive IoT can range from hundreds to billions, where the primary goal is to efficiently transmit a small amount of sensing data from a large number of devices.
Thus, establishing intelligent, efficient, adaptable, and cost-effective IoT systems in the context of a massive IoT paradigm is becoming a complicated task due to the increasing number of IoT connectivity demands and various application requirements \cite{9662390}. Therefore, connectivity becomes the core of IoT networks provided by various types of wired and wireless (terrestrial and non-terrestrial) communication technologies.

\begin{figure}[b!]
    \centering
    \includegraphics[width=0.98\linewidth]{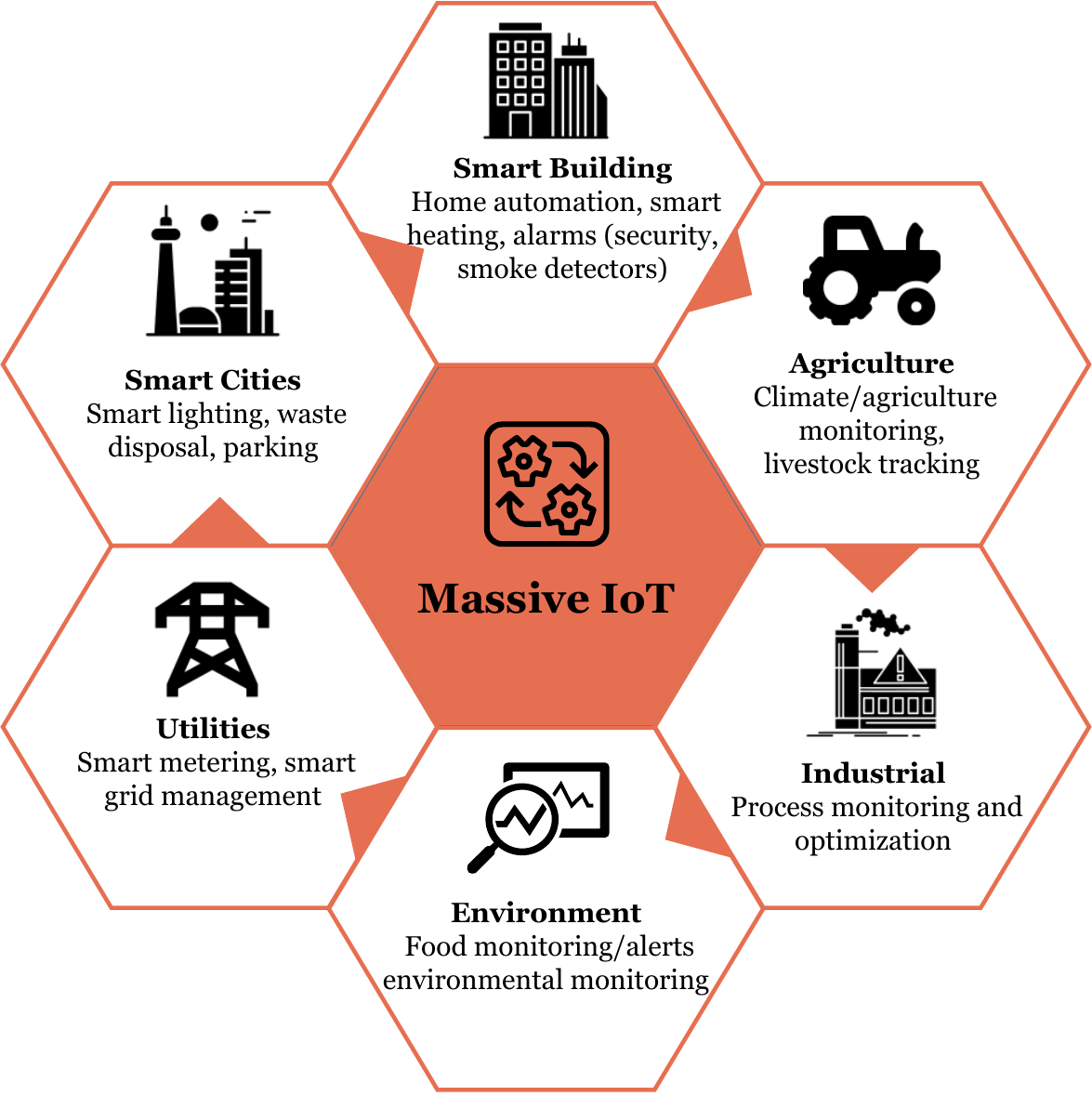}
    \caption{Massive IoT use cases enabled by LPWAN technology}
    \label{fig:massive-IoT}
\end{figure}

\begin{table*}[t!]
  \centering
  \renewcommand{\arraystretch}{1.8}
  \caption{Comparison of the most popular LPWAN technology types \cite{Yang2017, Hossain2021}}
  \begin{tabular}{l | p{11em} | p{10em} | p{11em} | p{11em} }
  \toprule
  &\textbf{LoRaWAN} & \textbf{SigFox} & \textbf{NB-IoT} & \textbf{LTE-M} \\
  \bottomrule
  \textbf{Technology} & Proprietary (PHY), Open (MAC) & Proprietary & Open LTE & Open LTE\\
    \hline
  \textbf{Spectrum} & Unlicensed & Unlicensed & Licensed & Licensed \\
  \hline
  \textbf{Frequency band} & Sub-GHz ISM &  Sub-GHz ISM & Cellular band & Cellular band \\
  \hline
  \textbf{Modulation} & CSS & D-BPSK & $\frac{\pi}{2}$-BPSK, $\frac{\pi}{2}$-QPSK  & BPSK, QPSK, 16\,QAM, 64\,QAM\\
  \hline
  \textbf{Duty cycle\/ TX restriction} & 1\% & 140 msg/day & Unlimited & Unlimited \\
  \hline
  \textbf{Frequency} & 433, 868, 915MHz & 868, 915MHz & 700-2100MHz & 700-2600MHz\\
  \hline
  \textbf{Bandwidth} & 125, 250, 500\,kHz & 100, 600\,Hz & 200\,kHz & 1.4\,MHz \\
  \hline
  \textbf{Coverage} & 1--10\,km & 10 -- 40\,km & 15\, km & 11\,km\\
  \hline
  \textbf{Battery life} &  10\,years &  10\,years &  10\,years &  10\,years \\
  \hline
  \textbf{Deployment} & Multi-operator, Self-deployment & -- & In band, Guard band LTE and GSM Standalone & In band LTE\\
  \hline
  \textbf{Standard} & LoRaWAN & No & 3GPP Release 13 & 3GPP Release 12\\
  \hline
  \textbf{Security} & AES-128 & AES-128 & LTE security & LTE security\\
  \hline
\end{tabular}
  \label{tab:LPWAN-technologies}
\end{table*}

Among different connectivity solutions, the low-power wide-area network (LPWAN) has been established as the preferred connection option for IoT networks due to its long communication range, low energy consumption, and low cost \cite{Furtado2020}. The LPWAN protocols can provide connectivity for many low-power battery-operated devices for delay-tolerant applications with limited throughput per device  \cite{10.1145/3544561}. Due to these attributes of LPWAN, IoT-based applications, machine-to-machine networks, and wireless sensor networks (WSNs) are projected to use LPWAN. Fig.~\ref{fig:massive-IoT} presents some massive IoT use cases enabled by LPWAN technology. The LPWAN protocol uses binary phase-shift keying (BPSK) and chirp spread spectrum (CSS) as physical (PHY) layer modulation schemes that tolerate interference and propagation noise to permit long-distance transmission \cite{Ma2021, jouhari2017implementation, Ouyang2017}. In addition, LPWAN nodes generally use low-quality and low-cost circuit configurations to meet the low-cost requirements  \cite{Raza2017}. In addition, LPWAN employs adaptive asynchronous random-access protocols on the medium access control (MAC) layer, such as the pure Aloha protocol with an additional acknowledgment mechanism due to its efficiency and simplicity as mentioned in LoRaWAN v1.0 specification \cite{alliance2015lorawan}.


Several LPWAN technologies have their own set of technical features, business models, and deployment strategies  \cite{10.1145/3300061.3345444}. We can classify LPWAN technologies into two categories based on the spectrum. First, narrowband IoT (NB-IoT) and long-term evolution for machines (LTE-M) standardized by the third-generation partnership project (3GPP) operate on the licensed spectrum. These types of technology provide high data rates (DRs) and bandwidth (BW) and ensure quality-of-service (QoS) at the cost of deploying complex protocols associated with high energy consumption and increased costs \cite{7552737, 7880946}. Second, SigFox and LoRaWAN standards operate on the unlicensed spectrum and compete to build their networks as quickly as possible. To enhance resistance against interference from other systems, devices in this latter category send a small signal over a larger frequency spectrum, enabling long-range communication with low power consumption. However, this strategy wastes the common spectrum and often encounters significant self-interference problems, limiting the total network capacity \cite{8660398}.

The above LPWAN technologies, including NB-IoT and LTE-M, which are based on LTE, enhance cell capacity and extend the device range to transmit or receive a small amount of data over a small BW. However, the conventional LTE-based solution is costly regarding the high power consumption of devices, complex protocols used to manage their functionality, and high deployment costs \cite{Amjad2021}. The IoT systems targeting LTE-M to preserve battery power due to the low power consumption of the LTE-M nodes, increase battery life by up to 10 years. Devices in the LTE-M release operate with 1.08~MHz BW, equivalent to six LTE physical resource blocks. Additionally, LTE-M can coexist with LTE services within the same BW, deploying different features to extend coverage, such as TTI bundling, repetition, and narrowband retuning. By reducing the complexity of IoT end devices (EDs), LTE-M has a low cost compared to second-, third-, and fourth-generation (2G, 3G, 4G) technology \cite{Shahgholi2021}. The other LTE-based LPWAN technology is the NB-IoT, a popular technology developed by 3GPP as part of Release 13. Although it is part of the LTE standard, it has a novel air interface, maintaining the essential requirements to reduce devices' costs and limit battery usage. Hence, the NB-IoT lacks several capabilities of LTE, such as handover, channel quality maintenance, carrier aggregation, dual connection, and other features. The NB-IoT operates under in-band, guard-band, and stand-alone modes. The stand-alone mode uses a GSM frequency with a BW of 200~kHz between guard bands of 10~kHz. The new guard-band and LTE resource block are used for the guard-band and in-band operation modes. The NB-IoT primarily aims to improve indoor coverage and enable connectivity between many low-throughput devices, achieving low latency and excellent service quality \cite{Ballerini2020}.

\begin{table}[t]
  \centering
  \renewcommand{\arraystretch}{1.2}
  \caption{Abbreviations and variables in this paper}
  \begin{tabular}{l l}
  \toprule
  \textbf{Abbreviations} &\textbf{Meaning}\\
  \bottomrule
  ABP & Activation By Personalization\\
  ADR & Adaptive Data Rate \\
  AES & Advanced Encryption Standard\\
  AoA & Angle of Arrival\\
  AppSKey & Application Session Key\\
  BLE & Bluetooth Low Energy\\
  BPSK & Binary Phase-Shift Keying\\
  BW & Bandwidth\\
  CDF & Cumulative Distribution Function \\
 CF & Carrier Frequency\\
 CR & Code Rate \\
  CRC & Cyclic Redundancy Check\\
  CSMA & Channel Sensing Multiple Access\\
  CSS & Chirp Spread Spectrum \\
  DR & Data Rate\\
  ED & End Device \\
  FEC & Forward Error Correction\\
  FFT & Fast Fourier Transform\\
  FSK & Frequency Shift Keying\\
  GW & Gateway \\
  ISM & Industrial, Scientific, and Medical\\
  JS & Join Server\\
  LEO & Low Earth Orbit\\
  LoRaWAN & Long Range Wide Area Network\\
  LoS & Line of Sight\\
  LPWAN & Low-Power Wide-Area Network\\
  mLBT & Multiple Listen-Before-Talk\\
  NLoS & Non-line of sight\\
  NOMA & Non-orthogonal Multiple Access\\
  $NF$ & Noise Figure\\ 
  NS & Network Server\\
  NwkSKeys & Network Session Keys\\  
  OSI & Open Systems Interconnection\\
  OTAA & Over-The-Air Activation\\
  PSLV & Polar Satellite Launch Vehicle\\
  $R_b$ & Bit Rate\\
  $R_c$ & Chirp Rate \\
  $R_s$ & Symbol Rate\\ 
  RSSI & Received Signal Strength Indicator\\
  RX & Receiving\\
  S-Aloha & Slotted Aloha \\
  SF & Spreading Factor\\
  SIC & Successive Interference Cancellation\\ 
  SIR & Signal-to-Interference Ratio\\
  SNR & Signal to Noise Ratio\\
  S-ALOHA & Slotted ALOHA \\
  TDoA & Time Difference of Arrival\\
  TP & Transmission Power\\
  TX & Transmission \\
  UAV & Unmanned aerial vehicle \\
  \hline
\end{tabular}
  \label{tab:abbreviations}
\end{table}
The second type of LPWAN technology that operates in the unlicensed band includes SigFox and LoRa technology. SigFox uses an industrial, scientific, and medical (ISM) spectrum of a 20~kHz BW with 868 or 915~MHz as the central frequency depending on the geographical area. 
The ISM band usage implies applying a duty--cycle restriction to LoRa EDs that restrict their uplink TX time. This duty cycle could be 0.1\%, 1\%, or 10\%, depending on the frequency band used for TX and on the local regulations. The research community in this field recommended a duty cycle of less than 1\% for efficient LoRaWANs \cite{9667513}, corresponding to the duty--cycle restrictions imposed by the European community. The restriction imposed for using ISM bands implies limiting the TX time for EDs. 
\begin{figure}[t]
    \centering
    \includegraphics[width=0.98\linewidth]{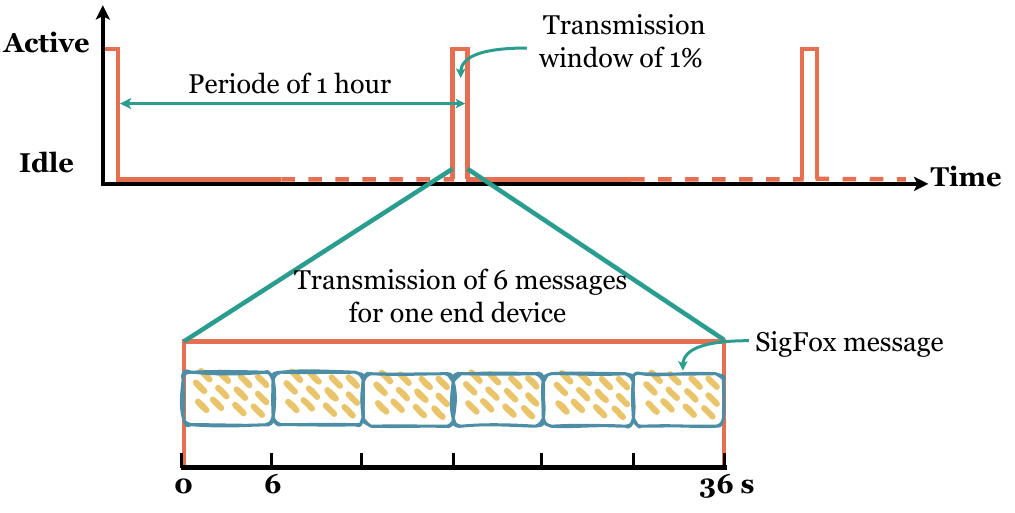}
    \caption{Effect of the 1\% duty cycle on the transmission (TX) time of an end device, considering the estimated TX time of SigFox messages of 6s.}
    \label{fig:dutycycle}
\end{figure}

For instance, the 1\% duty cycle regulation set by SigFox allows each ED to access the channel for only 36s within an hour. Moreover, knowing that the transmission period of a SigFox message is  6s, this allows each ED to send up to six messages an hour  (as illustrated in Fig.~\ref{fig:dutycycle}), for a total of 144 messages per day (24 hours $\times$ 6 messages) \cite{sigfoxQualificationSigfox}. However, due to the four messages reserved for protocol purposes by SigFox, EDs are only allowed to send 140 messages per day \cite{ShanmugaSundaram2020, Stusek2020}. The four reserved daily messages are used for various network management tasks, such as device registration, de-registration, network configuration, and network diagnostics. This 1\% duty cycle restriction limits the time EDs can use the channel to transmit messages, ensuring fair use of network resources, preventing network congestion, and ensuring that all devices have an equal opportunity to transmit their data. Nevertheless,  the 1\% duty cycle regulation can lead to an increased latency because the EDs must wait for the duty cycle window to open before sending data.

Compared to LoRa, SigFox offers an extended TX range of 10 to 40~Km versus 1 to 10~Km for LoRa networks \cite{10.1145/3293534, 10.1145/3345768.3355934}, allowing IoT devices to transmit collected data directly to a distant SigFox server. Moreover, SigFox technology is proprietary; therefore, its services are only offered to limited countries where a SigFox operator is established (75 countries worldwide)  \cite{sigfox-story, Haxhibeqiri2018}.

Due to its easy and low-cost deployment and remote maintenance requirements, LoRa technology has been identified as the most suitable LPWAN technology to ensure wireless connectivity for massive IoT networks. Moreover, LoRa is known for its worldwide availability due to its use of the specific sub-ISM band for communication, which helps deploy private networks without engaging external operators. Thus, the service provider will install and operate all the network equipment. In this scenario, a relay entity may be in charge of numerous IoT devices deployed around it to sense and collect data from the target area. Nevertheless, sporadic traffic pattern congestion and collision may arise since multiple IoT devices may simultaneously attempt to access the network and transmit data  \cite{10.1145/3356250.3361956}. This survey discusses the recent advances and challenges of a scalable LoRa network for massive IoT. More details are presented in the subsequent sections. Refer to Table \ref{tab:LPWAN-technologies} for a complete comparison between the mentioned LPWAN technologies. Table~\ref{tab:abbreviations} lists abbreviations and some variables used in this paper.

\subsection{Related Surveys}

\begin{table*}[]
  \centering
  \renewcommand{\arraystretch}{1.8}
  \caption{A summary of the most relevant related surveys}
  \begin{tabular}{c p{10em} c c p{35em} }
  \toprule
  \textbf{Surveys} & \textbf{Technologies} & \textbf{Nb Ref} & \textbf{Year} & \textbf{Short Description}\\
  \bottomrule
  \cite{Sinha2017} & LoRa, NB-IoT & 15 & 2017 & Provide a comparison between LoRa and the NB-IoT in terms of physical features\, network architecture\, and MAC protocol.\\
  \hline
  \cite{Raza2017} & SIGFOX, LoRa, INGENU RPMA, TELENSA, QOWISIO & 94 & 2017 & Surveys the design and techniques employed by different LPWAN technologies to offer wide-area coverage to low-power devices \\
  \hline
  \cite{UniversitateaTehnicaGh.AsachiIasi.FacultyofElectronics} & LoRa & 22 & 2017 & Evaluate LoRa modulation while considering the IoT requirement\\
  \hline
  \cite{Migabo2017} & LoRa, NB-IoT & 18 & 2017 & Survey the current research and provide a comparative study between LoRa and NB-IoT in terms of energy efficiency, reliability, and coverage\\
  \hline
  \cite{Saari} & LoRa & 40 & 2018 & Evaluate the usefulness of LoRaWAN technology in the field of IoT \\
  \hline
  \cite{Haxhibeqiri2018} & LoRa & 132 & 2018 & Provides an overview of published papers in IEEE explore database between 2015 and 2018, related to security, Physical and MAC layer\\
  \hline
  \cite{Sarker2019} & LoRa & 58 & 2019 & Explore different applications of LoRa and design a solution integrating Edge computing to enhance the performance of IoT-based applications\\
  \hline
  \cite{Qin2019} & LoRa, NB-IoT & 15 & 2019 & Provides a brief overview of LPWAN enabling technologies such as NB-IoT and LoRa\\
  \hline
  \cite{Ayoub2019} & LoRa, DASH7, SigFox, Wi-SUN & 68 & 2019 & Addresses the mobility of things and their connectivity in different LPWAN technologies\\
  \hline
  \cite{Erturk2019} & LoRa & 157 & 2019 & Gives a general introduction to LoRaWAN technology by discussing issues related to architecture and MAC protocols and provides open research opportunities.\\
  \hline
  \cite{Noura2020} & LoRaWAN & 205 & 2020 & Provides a survey on LoRaWAN security and privacy vulnerabilities\\
  \hline
  \cite{Alenezi2020a} & LoRa/LoRaWAN & 39 & 2020 & Reviews the challenges facing LoRaWAN in ultra-dense network \\
  \hline
  \cite{Kufakunesu2020} & LoRa/LoRaWAN & 52 & 2020 & Reviews the recent research on Adaptive Data Rate algorithms for LoRaWAN technology\\
  \hline
  \cite{Shahjalal2020} & LoRa & 15 & 2020 & An overview of LoRa-Based smart home using artificial Intelligence \\
  \hline
  \cite{Staikopoulos2020} & LoRa & 27 & 2020 &  Aims to highlight the available LoRa-based methods in the literature used to transfer images\\
  \hline
  \cite{M.Ortiz2020} & LoRa & 39 & 2020 & Analyses the performance of LoRa technology applied for vehicular communication using both experimental and simulation results\\
  \hline
  \cite{ShanmugaSundaram2020} & LoRa / LoRaWAN & 137 & 2020 & Surveys briefly LoRa-related issues and their recent solutions as energy consumption, communication range, error correction, and multiple access\\
  \hline
  \cite{Osorio2020} & LoRaWAN & 15 & 2020 & Provides an overview of LoRa-based multi-hop communication networks and summarizes different routing protocols designed to this end\\
  \hline
  \cite{Marais2020} & LoRaWAN & 73 & 2020 & Surveys the recent application of LoRaWAN that require confirmed traffic\\
  \hline
  \cite{DantasSilva2021} & LoRaWAN & 139 & 2021 & Presents a systematic review of the existing LoRaWAN optimization solution that aims to improve the IoT networking performance\\
  \hline
  \cite{Benkahla2021} & LoRa/LoRaWAN & 31 & 2021 & Provides an extended review and experimental evaluation of the latest ADR for LoRaWAN network\\
  \hline
  \cite{Ghazali2021} & LoRa / LoRaWAN & 89 & 2021 & Reviews the current research focused on UAV-based real-time LoRa communication\\
  \hline
  \cite{Boquet2021} & LoRa & 15 & 2021 & Provides a reference explanation of Long Range-Frequency Hopping Spread Spectrum (LR-FHSS)\\
  \hline
  \cite{electronics11010164} & LoRaWAN & 119 & 2022 &  Discuss the common features distinguishing different LPWAN technologies and defining their employability regarding the target scenario. \\
  \hline
  \cite{9667513} & LoRaWAN & 183 & 2022 & Gives a general insight into the energy efficiency in LoRaWAN networks, proposing some research directions in this context.    \\
  \hline
   Our Survey & LoRa/LoRaWAN & 274 & 2022 & We provide a survey on scalability challenges and the proposed solutions to assist LoRaWAN deployment in massive IoT networks\\
  \hline
\end{tabular}
  \label{tab:1}
\end{table*}

Due to the importance of LoRaWAN for providing connectivity to low-power devices, some specific surveys are available that focus on various aspects. This section covers multiple surveys on LoRaWAN and identifies the niche area of collision avoidance and interference management in LoRaWANs.
For instance, the authors of \cite{Noura2020} reviewed LoRaWAN architecture by describing the main components, such as EDs, GWs, networks, and application servers.
In addition, they \cite{Noura2020} presented a classification of the existing LoRaWAN attacks, such as the authentication, availability, and integrity attacks, and discussed some of the recent countermeasures to face the vulnerabilities mentioned above. Besides this, the drawbacks of LoRaWAN v1.1, released in 2017 to overcome these vulnerabilities, are discussed in detail. Most discussed vulnerabilities in this study are surpassed in the new LoRaWAN specifications. Alfonso et al. presented a brief overview of LoRaWAN routing protocols and the challenges of implementing multi-hop communication from a routing viewpoint \cite{Osorio2020}. However, the impact of multi-hop communication on the PHY layers and energy consumption of LoRa EDs were not covered in this study. The authors of \cite{Erturk2019} provided a general introduction to LPWAN technologies, especially the most popular type, which is the LoRaWAN, and compared SigFox, NB-IoT, LoRaWAN, and the conventional networks in terms of modulation schemes, frequency, coverage, BW, and energy consumption. In \cite{Erturk2019}, the authors also analyzed the LoRaWAN architecture using the layered open systems interconnection (OSI) reference model where LoRa is identified as the PHY layer and LoRaWAN as the MAC layer protocols. The authors of \cite{Staikopoulos2020} surveyed the most popular types of LPWAN technologies such as SigFox, DASH7, and LoRa. They summarized the general advantages and limitations of LoRa when used for image TX. Recently, the authors of \cite{Shahjalal2020} provided a brief overview of using machine learning with LoRa to enable remote smart-home monitoring.

The authors of \cite{Boquet2021} detailed the working process of the long-range-frequency hopping spread spectrum (LR-FHSS) as an extension of the LoRa physical layer. A comparison between the most popular types of LPWAN technologies (LoRaWAN, NB-IoT, and DASH7) was provided in \cite{Ayoub2019}, discussing how the mobility of devices can be handled by these technologies, generally considering fixed interconnected devices. Besides LoRa, more technologies enabling LPWAN are identified in \cite{Ayoub2019}, such as DASH7, SigFox, and Wi-SUN, based on an unlicensed frequency band. 
Recently, unmanned aerial vehicle (UAV)-based LoRa communication network deployment was reviewed in \cite{Ghazali2021}, where a UAV can be deployed in this architecture as an ordinary LoRa node or LoRa GW targeting either communication or localization applications. 
In this case, the wireless air-to-ground channel and UAV altitude play an essential role in the tuning process of LoRa PHY parameters for efficient deployment. Moreover, modeling the interaction between flying GWs is a challenging task due to their mobility and resource constraints. 
Mohammed et al. \cite{Alenezi2020a} provided an initiation of the possible challenges that may limit the deployment of LoRaWAN in ultra-dense networks and surveyed the existing software and hardware tools to simulate this scenario. However, none of the proposed solutions to assist the deployment of LoRaWAN in ultra-dense networks were discussed.

\begin{figure*}[t!]
    \centering
    \includegraphics[width=\linewidth]{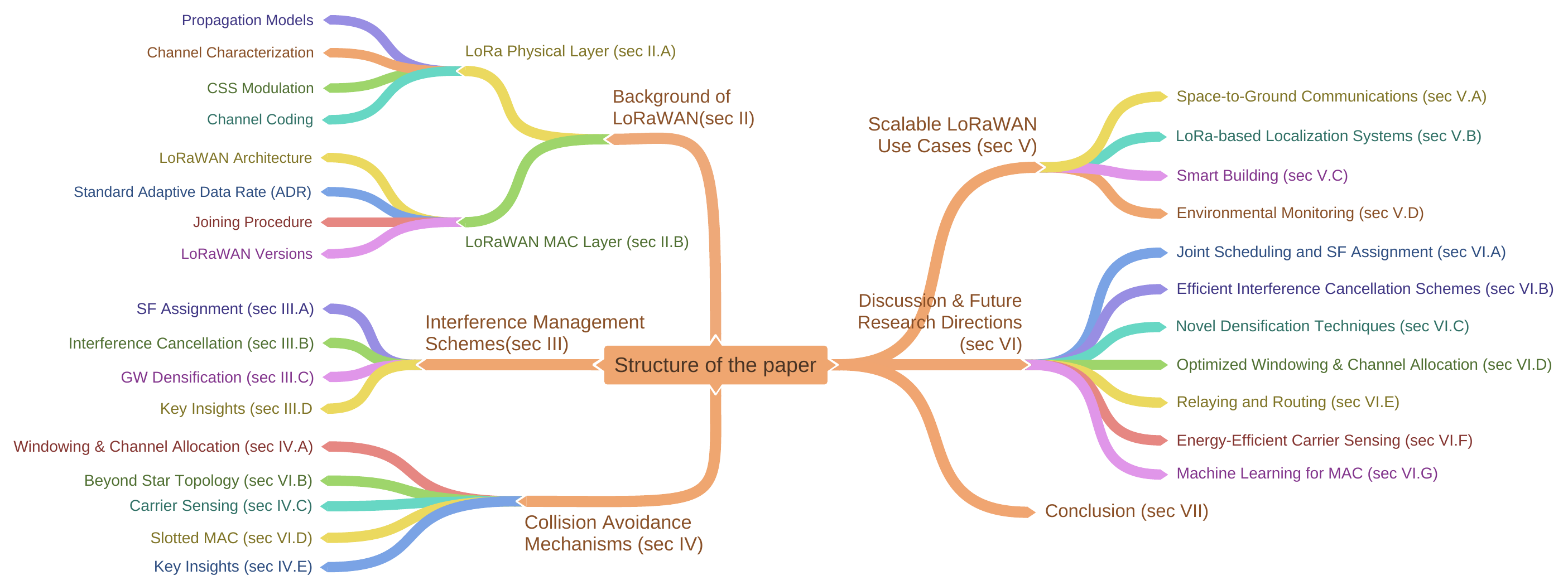}
    \caption{Structure of the paper}
    \label{fig:Structureofthepaper}
\end{figure*}

The LoRaWAN can self-adapt its configuration to enhance its performance in terms of reliability, radio range, delay, throughput, and power consumption. This adaptation is performed using various DRs allowed by the adaptive data rate (ADR) mechanism according to the following parameters: spreading factor (SF), BW, code rate (CR), and transmission power (TP) \cite{7996380}. The authors of \cite{Benkahla2021} provided a survey on different ADR enhancements existing in the literature and compared their performance in a mobile node scenario. Further, the authors of \cite{Kufakunesu2020} surveyed the optimization solutions for ADR. From a network perspective, the authors of \cite{9463416} distinguished five-factor categories that should be considered to evaluate the capability of LoRa to serve a crowded deployment efficiently:  PHY layer characteristics, deployment and hardware features, ED TX settings, LoRa MAC protocols, and application requirements. These parameters, defined as performance determinants were reviewed in \cite{9463416}. A systematic review of the existing LoRaWAN optimization solutions provided in \cite{DantasSilva2021, jouhari2023deep}, includes the co-existence of IoT devices and applications, resource allocation mechanisms, MAC layer protocols, network planning, and mobility issues.

The LoRa use case in vehicular communications was evaluated through a comparative study of the experimental and simulation results in \cite{M.Ortiz2020}. However, this study is performed from a PHY layer perspective making it unrealistic and incomplete. This outcome is due to the LoRa vehicular communication requirement for a sophisticated and customized LoRaWAN protocol to handle LoRa nodes' mobility and enable efficient data collection. The authors of \cite{Qin2019} provided a brief overview of the most popular LPWAN technologies such as LoRa and NB-IoT. This work focuses on explaining the useful insight for these types of LPWAN technology for efficient large-scale deployment based on measurements from a single deployment. This focus may limit the discussion this work provides to the specific use case. In \cite{Sarker2019}, the authors reviewed various LoRa applications to propose its integration with edge computing for the IoT  \cite{10.1145/3419016.3431491}. These applications were later obtained from the simulation in the discrete-event network simulator (NS-3) based on three metrics: packet delivery ratio, packet inter-reception time, and received signal strength indicator (RSSI). A survey on LoRa networking problems and solutions to enable its practical deployment were provided in \cite{ShanmugaSundaram2020}.

The review in \cite{Marais2020} studied LoRaWAN use cases that require confirmed traffic and discussed the viability of the actual traffic for the infrequent data transfer scenario. A general survey of LoRaWAN technology and the application fields was presented in \cite{Haxhibeqiri2018}, which distinguished between the PHY layer performance and network-level performance. Another generalized survey of LoRa technology is presented in \cite{Raza2017} (published in 2017), focusing on the performance of various types of technology that enable the LPWAN. The authors of \cite{Sinha2017, Migabo2017} provided a comparative study of LoRa and NB-IoT technologies, finding that based on the unlicensed spectrum, LoRa technology advances the NB-IoT in terms of battery life, capacity, and cost. The comparison would be more realistic if it were performed in the context of a specific LoRa use case to consider its fundamental requirements. In contrast, the licensed NB-IoT advances LoRa in QoS, latency, reliability, and range. Table~\ref{tab:1} compares the above surveys with this work.

  We can classify these surveys into four categories. The first class of surveys provides a comparative study of the different LPWAN technologies taking into consideration only the common features and performance metrics, such as the TX range and TP \cite{Raza2017, Sinha2017, Migabo2017,Qin2019, Ayoub2019, Staikopoulos2020}. However, these studies ignore the SF and the ADR in LoRaWAN, which requires further investigation. Moreover, the target LPWAN use-case requirements in terms of the deployment cost, TX range, and power consumption should be considered for an efficient comparative study of LPWAN technologies. Furthermore, the use-case location should be considered for a fair comparison due to its impact on the duty--cycle restriction applied to LoRaWAN communication and maximum TP, frequency plans, and BW. The second category of surveys in  \cite{Haxhibeqiri2018, Sarker2019,Shahjalal2020, M.Ortiz2020, Marais2020, Benkahla2021, Ghazali2021, 10.1145/3419016.3431491} focused on the application fields of the LoRaWAN in its minimum capacity considering simple use cases. The third category of surveys in \cite{ Erturk2019, Noura2020, DantasSilva2021, Boquet2021, 7996380} analyzed the LoRaWAN architecture using the layered OSI reference model or by describing the functionalities of each component. The fourth category of surveys in \cite{ ShanmugaSundaram2020, Alenezi2020a, Kufakunesu2020, Osorio2020, 9463416} addressed problems facing the efficient deployment of LoRaWAN in terms of networking, security \cite{benaddi2022adversarial, benaddi2021securing, jebari2022analysis}, and mobility. However, these studies were based on outdated LoRaWAN specifications. None of these surveys investigated the real challenges of recent studies facing the development of the LoRaWAN concerning its scalability due to its deployment in massive IoT. Unlike the mentioned surveys, our survey provides a deep study of the scalability issue and the latest solutions proposed in the literature for efficient LoRaWAN deployment in the massive IoT.

\subsection{Motivation}
To the best of our knowledge, recent state-of-the-art surveys published in the literature regarding the LoRaWAN address the challenges and open issues concerning energy efficiency and architecture. An essential part of these existing surveys focuses on comparing multiple types of LPWAN technology in terms of reliability, coverage, and mobility support. However, no current survey discusses the LoRaWAN scalability issues concerning massive IoT deployment despite its importance in numerous IoT applications, which presents a critical knowledge gap in this research area. The high connectivity demand is rising to become the primary problem facing the development of LoRaWANs regarding the number of proposed solutions that recently address LoRaWAN scalability, and the growing number of IoT devices that are expected to reach 22 billion by 2025 \cite{Zhang2020}. To address this problem, researchers in this field have suggested numerous solutions in the context of the PHY or MAC layer using the existing LoRaWAN features. Besides this, LoRa Alliance has released multiple specifications to address the problem of earlier versions of the LoRaWAN to enhance its functionality. Therefore, this paper presents this knowledge gap and its possible solutions for the LoRaWAN by providing an overview of the scalability issues and proposed solutions to improve the efficiency of LoRaWANs.

\subsection{Research Approach}
We conducted a comprehensive literature review to select, evaluate and analyze all relevant primary studies related to scalable LoRaWAN deployment in the massive IoT. We followed the systematic literature review approach \cite{kitchenham2007guidelines} for research paper inclusion in this survey. In the first step, we classified the articles, then we analyzed and differentiated the selected papers through deep reading and content analysis. Selected papers are classified into the following five main categories: 
\begin{enumerate}
    \item Surveys:  All relevant review papers focused on LoRa communication system issues providing a comparative study of LPWAN technologies.
    \item Background: Recent papers discussing LoRaWAN capabilities, describing LoRa features, and analyzing CSS modulation-based LoRa communication. 
    \item PHY layer:  Papers addressing the LoRaWAN scalability issue from the PHY layer perspective.
    \item MAC layer: Papers contributing to LoRaWAN scalability by improving LoRaWAN MAC layer protocols.
    \item Use cases: Recent papers on LoRa-based massive IoT applications requiring scalable LoRaWAN.
\end{enumerate}

\subsection{Contributions}
Unlike the mentioned surveys that focus on comparing different types of LPWAN technology or providing a general review of LoRa technology (where security issues and energy efficiency are the main interests), this work reviews the proposed solutions in the literature to address the LoRaWAN scalability issues facing its deployment in the massive IoT. The contributions of this work are summarized as follows:

\begin{itemize}
   \item First, we review the LoRaWAN system and its features from the perspective of the PHY and MAC layers. 
    \item Next, we discuss the issue of LoRa signal interference when several EDs transmit their signals simultaneously using the same combination of LoRa PHY layer parameters. Moreover, we present different solutions proposed in the literature to overcome this issue and classify them into three categories: SF assignment, interference cancellation, and GW densification.
    \item Then, we investigate the existing collision avoidance schemes in the literature to address the collided LoRa concurrent TX at the MAC layer level. We also categorized the current solutions such as windowing and channel allocation, beyond the start topology, carrier sensing, and slotted MAC protocols. 
    \item Afterward, we present several use cases and applications of the LoRaWAN that may involve the massive IoT such as space-to-ground communications, LoRa-based localization systems, smart buildings, and environmental monitoring.
    \item Finally, we discuss the proposed solution drawbacks, remaining open issues, and potential future research directions for LoRaWAN deployment in ultra-dense IoT networks.
\end{itemize}

\subsection{Organization}
The rest of the paper is organized as follows: Section \ref{sec:LoRaWAN}, describes the LoRa PHY layer specifications, including LoRa modulation, SFs, and propagation models. Moreover, Section~\ref{sec:LoRaWAN} presents the LoRaWAN architecture as well as the ED classes and joining procedure. Schemes to manage and overcome LoRa signal interference are discussed in Section~\ref{sec:interference}. Section~\ref{sec:Collision} outlines the current collision avoidance mechanisms to allow concurrent LoRa TX in dense deployments. The latest use cases and applications of LoRaWAN are classified and reviewed in Section~\ref{sec:Applications}. Finally, we conclude the paper in Section~\ref{sec:Future} by discussing the advantage and disadvantages of the proposed solutions to enhance the capacity of LoRaWAN, allow its deployment in dense networks, and identify the remaining open questions in this field. Fig.~\ref{fig:Structureofthepaper} provides a general overview of the paper structure.

\section{Background of LoRaWAN}\label{sec:LoRaWAN}
This section describes the specification of LoRa technology that influences its capacity to serve highly dense networks as massive IoT networks. It presents the first step to understanding the challenges facing the deployment of LoRa in a massive IoT network.
\subsection{LoRa Physical Layer}\label{LoRaphysical}
The LoRa network relies on two key components, LoRa and LoRaWAN, where LoRa presents the physical layer modulation whereas the LoRaWAN represents the MAC layer protocol. Combining these two components results in a low-power and cost-effective wide-area network. We discuss the PHY and MAC layer issues separately in the following sections.

\subsubsection{Propagation Models}
A network planner's ability to effectively predict the LoRa signal behavior in the target environment is a critical part of efficient network deployment. It allows predicting the ED coverage to help determine the number of GWs required for a set of EDs. Propagation models are the most frequent method due to their ability to estimate the received signal strength based on path-loss calculation. Based on the derivation of the resulting path loss, wireless propagation models can be split into three categories: (i) empirical, (ii) deterministic, and (iii) stochastic; more details are provided in \cite{Stusek2020}. To date, a variety of propagation models for wireless technologies have been presented, with the majority of them coming from three key sources: (i) standards, (ii) vendors/operators, and (iii) academics.
 The Okumura-Hata model \cite{joseph2013urban} is commonly used as a propagation model in LoRaWAN. The Hata model \cite{joseph2013urban} provided mathematical equations that are correlated to the Okumura model for different parameters. These models aid in the analysis of propagation impact using computer simulations, where path loss is computed based on characteristics such as the frequency range (150 to 1500 MHz), distance range (1 to 20 km), and antenna height.

The COST 231-Hata-model \cite{4510943} is a variation of the Hata model, extending its frequency range. The path-loss prediction model is based on basic system characteristics, where the frequency varies from 1500 to 2000~MHz, the distance between the EDs and GWs ranges from 1 to 20~km, while the antenna height is 1 to 10~m and 30 to 200~m. COST-231 Walfish-Ikegami model referred to COST-WI \cite{hari2003channel}, is a hybrid of the Walfish and Ikegami models that enhance path loss prediction by taking into consideration more variables to define large and medium-sized urban environments \cite{7765635},  namely the road widths, building heights, building spacing. Other propagation models were evaluated in \cite{Harinda2019, 8739575, 8767299} for the LoRaWAN use cases in urban, indoor, and outdoor environments, which are not in the scope of this paper.

\subsubsection{Channel Characterization}
Establishing the signal-to-noise ratio (SNR) threshold $SNR_{th}$, which is based on how to compute the receiver sensitivity ($S$), is a critical component of LoRa wireless communication systems. There are different $SNR_{th}$ for each SF that is valid in the absence of interference for accurate demodulation of received signals. Table~\ref{tab:SNRth} lists the dependency of SNR limit on LoRa features, such as the SF, DR, and the number of chips per symbol. The noise-floor power level divided by the receiver sensitivity yields this threshold. For each SF provided by Semtech's report, the SNR threshold changes by 2. 5 dB for every increase in SF. The receiver sensitivity of a LoRa transceiver using the Bandwidth $BW$ is calculated as follows for each SF:
\begin{equation}
    S = -174 + 10.log_{10}(BW)+ NF + SNR_{th},
\end{equation}
where NF = 6 is the noise-figure tolerance at a GW and is fixed for a hardware implementation, whereas -174~dBm is the thermal noise density primarily affected by the receiver temperature.
\begin{table}
    \centering
    \renewcommand{\arraystretch}{1.8}
    \caption{Corresponding spreading factor for each chip/symbol, SNR limit, and data rate for LoRa communication over a channel of 125 kHz BW}
    \begin{tabular}{c c c c}
        \toprule
        \textbf{Spreading Factor} & \textbf{Chips/Symbol} & \textbf{SNR limit} & \textbf{Data Rate }\\
        \bottomrule
        7 & 128 & -7.5 db & 5469 bps\\
        \hline
        8 & 256 & -10 & 3125\\
        \hline
        9 & 512 &	-12.5 & 1758\\
        \hline
        0 & 1024 &-15	& 977\\
        \hline
        11 & 2048 & -17.5 & 537\\
        \hline
        12 & 4096 & -20	& 293\\
        \hline
    \end{tabular}
    \label{tab:SNRth}
\end{table}

The CSS modulation supports variable data TX rates using quasi-orthogonal SFs. As a result, depending on the deployment field, the system is designed to swap DRs to efficiently enhance its TX range or power consumption to optimize network performance over a constant BW. The symbol and the bit rates are given in the function of the BW at a specific SF. Thus, increasing the BW by a factor of two doubles the TX rate. The relationships between the symbol bit rate $R_s$, $BW$, $SF$, and the conventional bit rate $R_b$ are as follows:
\begin{equation}
    R_s = \frac{BW}{2^{SF}} 
\end{equation}
\begin{equation}
    R_b = BW*R_s*FEC,
\end{equation}
where $FEC$ is the LoRa modem's Forward Error Correction rate, which protects against symbol errors caused by interference. The chirp rate ($R_c$) can be calculated as a function of the symbol rate as follows:
\begin{equation}
    R_c = BW * R_s = \frac{BW^2}{2^{SF}}
\end{equation}

The FEC necessitates the insertion of error correction bits (redundant bits) into the transmitted data. Although the FEC reduces data throughput, it improves receiver sensitivity. Because the rejection gain fluctuates from 16 to 36 dB, signals from various SFs are considered periodically orthogonal. As a result, LoRa may allow up to six simultaneous TXs on a single channel using six different adjustable SFs ranging from 7 to 12. The LoRa signals are protected from interference by adapting the CR, and such a robust system uses greater CR. Similarly to the SF, a greater CR value indicates that a long packet can be transmitted using high energy consumption, and such a CR takes a value from \{4/5, 4/6, 4/7, 4/8\}. DR parameter in LoRa is calculated using the SF, BW, and CR as follows:
\begin{equation}
    DR = SF \times \frac{BW}{2^{SF}} \times CR.
\end{equation}

\begin{figure}[t]
    \centering
    \includegraphics[width=0.98\linewidth]{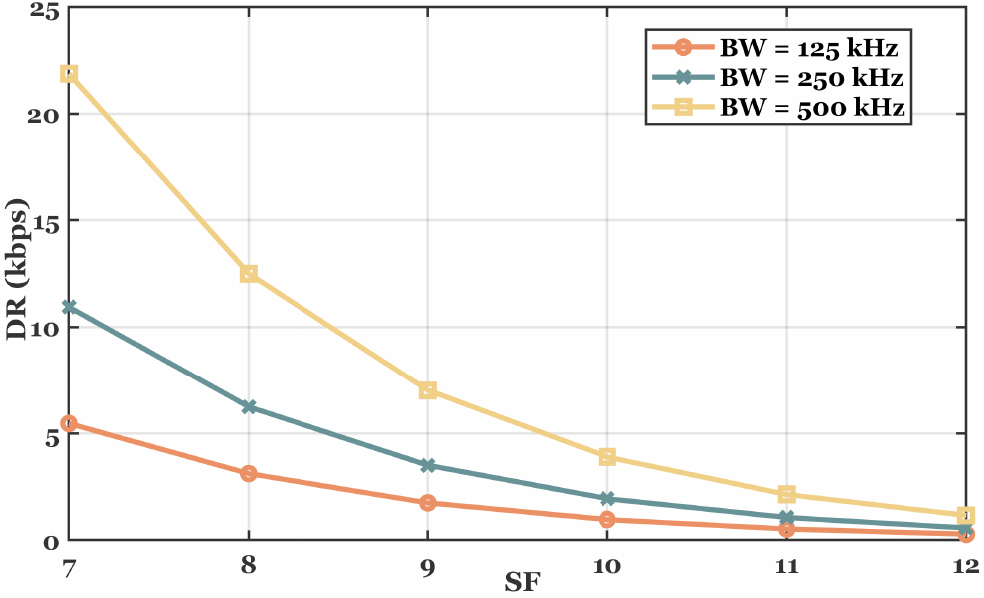}
    \caption{Data rate versus spreading factor for CR =4/5}
    \label{fig:Data_rate}
\end{figure}

Fig.~\ref{fig:Data_rate} plots the results of DR in the function of the SF and BW provided by Equation 5. The figure reveals that a high BW ensures an increased DR, whereas increasing the SF decreases the DR due to the long TX range of a high SF.

\begin{figure}[t]
    \centering
    \includegraphics[width=0.98\linewidth]{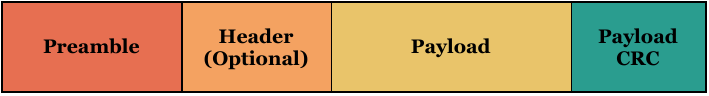}
    \caption{LoRa frame structure}
    \label{fig:Frame_structure}
\end{figure}

The preamble, optional header, the data payload, and optional cyclic redundancy check (CRC) field are all part of the PHY layer packet structure (Fig.~\ref{fig:Frame_structure}). The preamble is needed to synchronize the receiver with the transmitter, and it can have between 10 and 65,536 symbols in total. The first four symbols of the preamble are fixed, while the remaining are programmable, with a minimum of six symbols and a maximum of 65,532 symbols. The payload length, the payload CR, and the header CRC are all included in the optional header. The presence of the header is determined by the header mode (explicit or implicit), and it is deactivated in the implicit mode only when the payload length, CR, and CRC settings are known in advance or fixed. Consequently, the time on air of the packet decreases. In the explicit mode, the payload length in bytes, payload FEC coding rate, and the header CRC are all included in the header. The FEC is always used to secure the header, with the greatest coding rate of 4/8. The payload may contain either MAC layer information related to the LoRaWAN standard or data packets \cite{Wu2020}.

\subsubsection{CSS Modulation}
\begin{figure}[h] 
    \centering
  \subfloat[\label{fig:1a}]{%
       \includegraphics[width=0.98\linewidth]{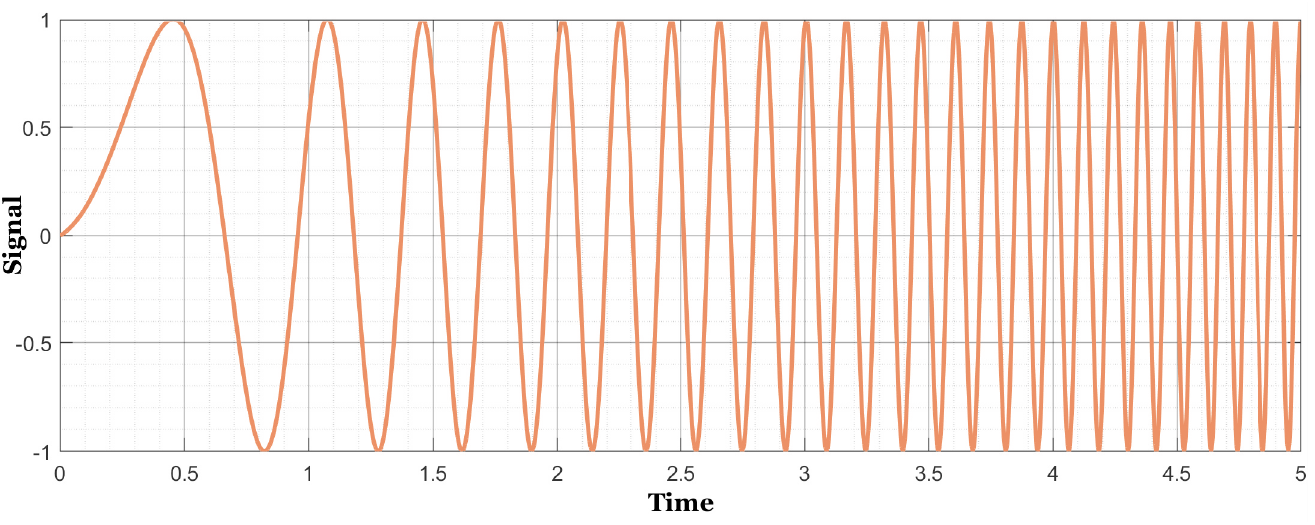}}
    \\
  \subfloat[\label{fig:2b}]{%
        \includegraphics[width=0.98\linewidth]{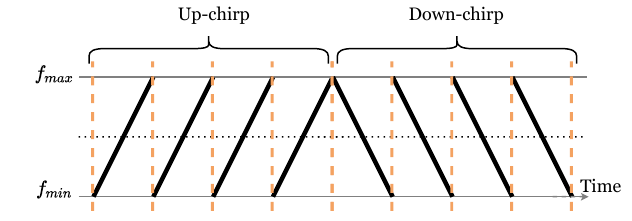}}
  \caption{LoRa chirp signal}
  \label{fig:Chirp_signal} 
\end{figure}

Semtech released the LoRa PHY layer in 2014 adopting a modulation variant of CSS \cite{Wu2021}. The CSS enables it to reach great distances while resisting interference, fading, and Doppler effects. Frequency shift keying (FSK) modulation may also be employed. Although FSK achieves low-power consumption, it falls short of LoRa CSS's communication range. The CSS modulates the information-carrying signal in the form of a series of chirp pulses and passes it through an FEC mechanism before the TX. A chirp is a temporal profile of the frequency response that shifts from one frequency $f_0$ to $f_1$ during time interval $T$ (Fig.~\ref{fig:1a}). We can distinguish two types of chirps in LoRa as indicated in \cite{Edward2021}. First where the frequency linearly increases from the minimal frequency $f_{min}=-\frac{BW}{2}$ to the maximal frequency $f_{max}=-\frac{BW}{2}$ is called the base chirp (Fig.~\ref{fig:2b}). The second type is the down-chirp, whose frequency shifts from $f_1$ to $f_2$, the complex conjugate of the base chirp.

In CSS,  a modulator generates multiple chirps for different digital inputs, each with a distinct temporal shift from the base chirp. In addition, CSS uses a sequential series of rising and falling chirp pulses to transmit data, such as chirp is a symbol, and the symbol time is the length of time between chirps. Unlike the direct sequence spread spectrum, CSS spreads frequencies using chirp pulses rather than pseudo-random code sequences. The number of bits encoded for a symbol is the SF (ranging from 7 to 12). For instance, if the SF is 7, it is feasible to send 7 bits and 256 distinct patterns over one symbol; if SF is 8, it is possible to send 8 bits and 512 different patterns. The initial frequency of the chirp indicates the values taken by the symbol. The SF can be calculated based on the number of chirps $N$ as follows: 
\begin{equation}
    SF = \log_2(N) = \log_2(\frac{R_c}{R_s}),
\end{equation}
where $N = \{128, 256, 512, 1024, 2048, 4096\}$,  and $R_c$, and $R_s$ are  the chirp and symbol rate, respectively.

Besides modulation, LoRa communication performance may be adjusted by modifying other PHY layer configuration parameters such as the TP, carrier frequency (CF), SF, BW, and CR. The BW can take different values from 125, 250, and 500~kHz, a higher value indicates that the packet may be sent with a high DR but for a shorter distance due to the low receiver sensitivity. A higher SF implies a lower SNR required to achieve reliable communication. Consequently, as the distance between the transmitter and receiver increases, the BW must be lowered and the SF increased to transmit longer packets at the cost of higher energy consumption (see Table \ref{tab:LoRa_parameters}). 
A typical LoRaWAN consists of LoRa EDs communicating data via uplink TX to multiple GWs that notify EDs using LoRa-based downlink messages after forwarding received data to the LoRaWAN server. The LoRa transceivers are equipped with four blocks to encode/decode LoRa signals exchanged with LoRaWAN GWs. These components are discussed in the following section. 

\begin{table}
    \centering
    \renewcommand{\arraystretch}{1.8}
    \caption{An overview of LoRa's transmission parameters and their impact on the link performance and energy consumption}
    \begin{tabular}{p{4em} p{7em} p{16em} }
        \toprule
        \textbf{Parameters} & \textbf{Values} & \textbf{Impacts} \\
        \bottomrule
        TP & -4 to 20~dBm & High transmission power implies a lower SNR, which will increase the energy consumption of the transmitter \\
        \hline
        CF & 137 to 1020~MHz & Transmissions that overlap in time, but not in carrier frequency, do not interfere with each other and can be decoded simultaneously \\
        \hline
        SF & $2^6\, to\, 2^{12} \frac{chips}{symbol}$  & A high SF increases the SNR, leading to an increased radio sensitivity and range, at the cost of longer packet airtime, resulting in higher energy consumption \\
        \hline
        BW & 125 to 500~kHz & A high BW implies a high data rate for packet transmission, leading to reduced radio sensitivity and communication range \\
        \hline
        CR & 4/5 to 4/8 & As CR increases, transmission becomes more resilient to interference bursts, resulting in longer packets and more energy consumption\\
        \hline
    \end{tabular}
    \label{tab:LoRa_parameters}
\end{table}

\subsubsection{Channel Coding}
As illustrated in the LoRa PHY layer block diagram, the LoRa transceiver encodes and decodes data bits in four steps. An encoder first encodes the binary source input bits. The LoRa uses Hamming codes for FEC, which are linear block codes that are easy to implement. The chosen CR value determines the encoder output. Encoding lowers packet error rates when short bursts of interference occur. As mentioned, the LoRa standard uses a CR of 4/5, 4/6, 4/7, and 4/8 such that for a code rate $k/n$ the number of useful information bits equals $k$, whereas $n$ is the output generated by the encoder and ($n - k$) is the number of redundant bits. The second phase minimizes the correlation of the redundant bit added by the Hamming encoder by deploying a whitening block. The whitening block introduces a known sequence that depends on the CR for the payload only \cite{10.1145/3293534}, enhancing the system resistance against frequency selective fading \cite{Courjault2021}. The third phase applies an interleaver of ($k \times r$) that transposes and shifts digits of symbols to map the encoded data into LoRa symbols. This process consists of transposing each $k$ codeword and shifting its digits by $r \pmod{SF} $ to the left. In the fourth phase, the errors of adjacent bits are reduced by applying gray coding to the interleaved symbols \cite{grcon}.

\begin{figure}[t]
    \centering
    \includegraphics[width=0.98\linewidth]{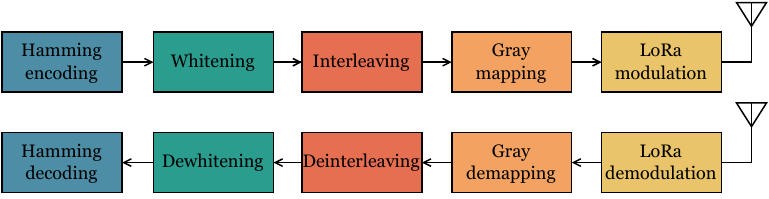}
    \caption{LoRa channel coding diagram}
    \label{fig:channel_coding}
\end{figure}

\subsection{LoRaWAN MAC Layer}
\begin{figure}[t]
    \centering
    \includegraphics[width=0.98\linewidth]{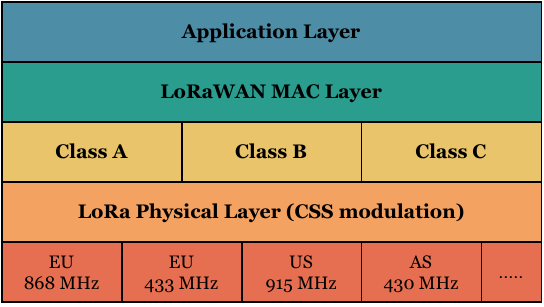}
    \caption{LoRa technology overview}
    \label{fig:LoRaWAN_architecture}
\end{figure}

The LoRaWAN (MAC) protocol is an open-sourced protocol built on the top of LoRa \cite{alliance2015lorawan} PHY layer and was standardized by LoRa Alliance Fig.~\ref{fig:LoRaWAN_architecture}. This protocol aims to manage the MAC mechanism to allow communication between numerous devices and the network GWs. Devices in LoRaWANs access the wireless channel randomly in the pure Aloha mode, as stated in the LoRaWAN standard, and must comply with a precise duty cycle. If the hidden node problem is considered, and the listen-before-talk phase is skipped for energy savings, Aloha MAC protocol usage is justified \cite{doi:10.1177/1550147717699412}.

\subsubsection{LoRaWAN Architecture}
\begin{figure*}[t]
    \centering
    \includegraphics[width=0.98\linewidth]{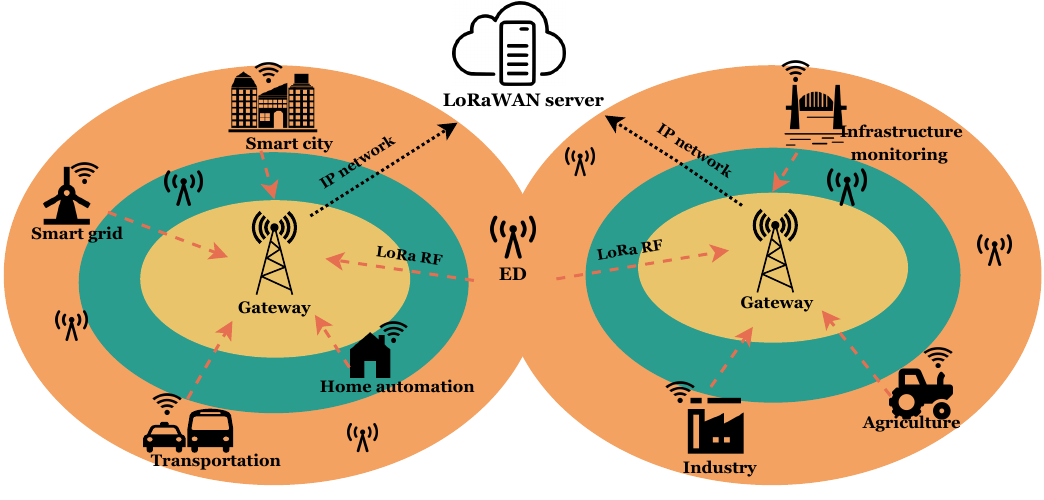}
    \caption{Legacy LoRaWAN architecture: EDs deployed in the target area in the context of any IoT use cases such as smart grid, smart city, or infrastructure monitoring. Each ED uses its LoRa link to transmit the collected data to the GW, whereas this latter forwards the aggregated data to the server via an IP network link.}
    \label{fig:LoRaWAN_application}
\end{figure*}

The LoRaWAN features in a star-of-star topology \cite{Centelles2021}, comprising three main components: network servers (NSs), GWs, and EDs. Thus, EDs may only connect with LoRaWAN GWs and not with each other directly as peer-to-peer (Fig.~\ref{fig:LoRaWAN_application}). A central NS is responsible for connecting all the GWs in the network. Only the LoRaWAN GWs can deliver collected data packets from end nodes to the NS, encapsulating them in UDP/IP packets. The NS can transmit downlink packets and MAC instructions to EDs if necessary. Furthermore, communication ends at application servers that may or may not be hosted by third parties. A specific NS can support many application levels. Fig.~\ref{fig:LoRaWAN_architecture} depicts the resulting LoRaWAN architecture. Although communication is bidirectional in LoRaWAN, it is strongly suggested to communicate over the uplink. The GW collects data from the nearest EDs within its communication range, thus a collision may occur when receiving multiple messages simultaneously over the same channel.

LoRaWAN defines three classes of EDs \cite{Donmez2018}:
\begin{itemize}
   \item Class A devices have fundamental features that every ED must have to join the LoRaWAN. In this class, each uplink TX is followed by two short receiving windows wherein the ED watches for the incoming downlink traffic to enable the bidirectional connection. Thus, the ED initiates downlink communication, pushing each downlink TX to wait for the duration of the uplink TX. The first and second downlink receive windows begin for 1 and 2~s, respectively after the uplink broadcast ends. The network server is in charge of scheduling downlink traffic and performing timing control. Because they sleep most of the time, Class A EDs use the least amount of energy \cite{Furtado2020}.
    \item Class B EDs provide extra receive windows to augment downlink possibilities at predetermined periods. The GWs send downlink beacons to synchronize Class B EDs and inform the NS when specific EDs listen for downlink communication. Class B devices require more power than Class A devices because they must open more receive windows, although these windows are not used for downlink communication.
    \item The Class C device has practically continuous open receive windows that close during broadcasting. Thus, Class C has the highest power consumption.
\end{itemize}

\subsubsection{Standard Adaptive Data Rate (ADR)}
The ADR is one of the essential features of LoRaWAN, which aims to reduce the energy consumed by EDs to transmit data while maximizing throughput by adjusting the DR (consequently, the SF as provided in Table~\ref{tab:SNRth}) and TP. The ADR can manage TX parameters at both the ED and NS sides \cite{sornin2015lorawan}. The server developer created the NS algorithm, whereas the LoRa Alliance established the ED algorithm. The ED is responsible for deciding whether the ADR should be employed. Once the NS is on, it regulates the device's TX parameters and transmits the specified ADR commands to the ED. In addition, the device should verify whether the NS still regularly receives its uplink frames; if not, it should adjust its SF. In a LoRaWAN, DR adjustment enables easy network scalability by deploying additional GWs. Moreover, the ADR may significantly enhance the network capacity since the SF factor management allows the TX of multiple packets simultaneously using different SF, which are orthogonal \cite{7996384}. The ADR mechanism can also deploy FSK modulation in some LoRaWAN use cases requiring a high data rate since the CSS-based LoRa modulation offers a DR ranging from 0.3 to 37.5~kbps. In comparison, FSK provides a higher DR of up to 50~kbps. The ADR mechanism was designed for the LoRaWAN to regulate the end node' TX settings to increase the packet delivery ratio. The TP is defined for the uplink (from the ED to the GW) based on the environmental conditions.

The ADR mechanism allows EDs to dynamically manage the DR and TP, relying on the link budget estimated in the uplink communication alongside the SNR threshold necessary for correctly decoding the received data packets at the allowed DR. The NS maintains the ADR for the fixed-end nodes based on previously received packets over the uplink communication, also known as "network-managed ADR" or "static ADR". In the case of mobile EDs, the NS ADR does not function properly due to the time variance of channel attenuation resulting from device movement. Thus, the ADR is performed on the side of the ED, called a "blind ADR" \cite{MOYSIADIS2021102388}. Moreover, ADR LoRaWAN may use adaptive modulation techniques with various communication channels and modem transceivers at GWs to simultaneously receive numerous messages from the channels. Every signal adopts a different SF, orthogonally separated as enabled by the spread spectrum \cite{FEHRI20181096}. The ADR performance is significantly reduced under a very crowded network and time-variance channel circumstances \cite{8647469, 8406255}. Furthermore, because EDs are mobile, topology changes often, affecting the link quality between EDs and a GW  \cite{10.1145/3349621.3355727}.

\subsubsection{Joining Procedure}
New EDs must undergo an activation procedure to join and authenticate themselves in the LoRaWAN. To this end, two-session keys are exchanged between the EDs and the NS throughout this activation procedure. The LoRaWAN protocol provides two modes of activation. The first one is over-the-air activation (OTAA), and the second is activation by personalization (ABP) \cite{sornin2016lorawan}. In OTAA, the joining procedure is performed between the EDs and the NS to participate in the data communication process. This process consists of exchanging messages of join request and join accept. The ED starts the process by sending the join request to the NS, consisting of the devEUI, AppEUI, and DevNonce. The LoRaWAN uses this latter to drop some joining requests where the DevNonce is similar. Thus, the system can identify some reply attacks. When the join server (JS) successfully receives the join request, a join accept message is sent to the ED, including the newly generated JoinNonce. Using the previous information from request message fields and Join-Nonce, the application session key (AppSKey) and network session keys (NwkSKeys) are created by the JS to enable device access to LoRaWAN services \cite{Choi2020}. The AppSKey encrypts data transmitted by the EDs, ensuring integrity using the NwkSKey. Those keys are dynamically assigned to the EDs at each join process, improving network security and introducing additional complexity in the join process. However, there is no need for joining in the activation by personalization process since the AppSKey and NwkSKey are fixed and pre-stored on the EDs. The AppSKey and NwkSKey are unchangeable regarding the activation session. This process reduces the join process complexity and the number of exchanged messages to access the network while weakening the security level of the LoRaWAN.

\subsubsection{LoRaWAN Versions}
Different versions of LoRaWAN have been released to enhance its performance in terms of security, scalability, and real-time long-range communication regarding the new advancement in resource-constrained IoT devices. The first specification, LoRaWAN v.1.0 (also known as the draft version) was released in January 2015 \cite{alliance2015lorawan} to describe the legacy LoRaWAN protocol and provide a detailed description of the classes. In addition, LoRa Alliance issued a few minor changes and added more clarification to LoRaWAN v1.0.1 released in February 2016 \cite{lorawanv101}. The main changes introduced in this version were the additional clarification of the RX window start time definition, the requirement for using a coding rate of 4/5 to guarantee a maximum time-on-air $<$ 400~ms, and the correction of the JoinAccept MIC (The message integrity code) calculation. Further, LoRaWAN v1.0.2 was released in July 2016 \cite{lorawanv102} and introduced two distinct keys NwkSKey and AppSKey, which are both derived from a single root key. With this specification, the PHY layer section describing the regional parameters was improved and moved to a separate document ("LoRaWAN Regional Parameters" \cite{lorawanv1022}). Furthermore, LoRaWAN v1.0.2 proposed countermeasures to address replay attacks by encrypting the uplink frame counter (FCntUp) and using it in the downlink confirmation messages (ACKdownlinks).

Moreover, LoRaWAN v1.1 was released in October 2017 \cite{lorawanv11} that included many drastic changes to face the security vulnerabilities in the previous versions (i.e., v1.0.1 and v1.0.2) resulting from weak authentication, weak key management, and weak encryption. Additionally, the release dealt with security attacks, such as ACK spoofing, eavesdropping, and packet modification \cite{Noura2020, Eldefrawy2019}. This is through enabling handover roaming alongside the passive roaming already supported in v1.0.x and enhancing the security of the OTAA  procedure by introducing the JS logical entity responsible for managing the ED authentication process. In v1.1, multiple downlink frame counters (FCntDown) are used instead of a single one shared over all ports. Furthermore, reset indication commands (ResetInd, and ResetConf) and rekey indication commands (RekeyInd, and RekeyConf) are introduced for the LoRaWAN v1.1 server, respectively, for ABP and OTAA devices.

In July 2018, LoRaWAN v1.0.3 was released to update specifications (i.e., for v1.0.1, v1.0.2, and v1.1) by importing Class B functionalities from v1.1 into v1.0.x branch (introducing unicast/multicast support to Class B devices). The DeviceTimeReq and DeviceTimeAns MAC commands for Class A devices were added and the BeaconTimingReq and BeaconTimingAns commands in v1.0.1 and v1.0.2 were removed. The DeviceTimeReq command allows the EDs to request the current network time from an NS to synchronize its internal clock and speed up the acquisition of beacons sent by Class B devices. The NS sends the current network time as a payload of the BeaconTimingAns command to answer the ED request. Recently, LoRaWAN v1.0.4 was released in October  2020, which is the latest protocol specifications for new LoRaWAN developments. However, v1.0.2 and v1.0.3 EDs still dominate the field due to their hard upgrade as they have different hardware requirements. The current specifications of LoRaWAN 1.0.4 \cite{lorawanv104} have made it possible to improve the security breaches identified in its previous specifications (LoRaWAN 1.0.2) to secure communications between terminals and the core network operators. The JS introduced during the LoRaWAN 1.1 (2017) specification for roaming management is also present in the LoRaWAN 1.0.4 (2020) specification. The LoRaWAN 1.0.x specifications authorize passive roaming, and the LoRaWAN 1.1 specification authorizes both passive roaming and handover roaming. The JS server allows greater flexibility in Roaming management and guarantees the independence of data encryption keys. Table \ref{tab:LoRa_versions} summarizes the evolution of the LoRaWAN specifications, including the year of release and major changes.


\begin{table}
    \centering
    \renewcommand{\arraystretch}{1.8}
    \caption{The evolution of LoRaWAN specifications}
    \begin{tabular}{p{3em} p{2em} p{16em} p{4em}}
        \toprule
        \textbf{Version} & \textbf{Year} & \textbf{Main changes} & \textbf{Reference}\\
        \bottomrule
        1.0 & 2015 & Draft version & \cite{alliance2015lorawan}\\
        \hline
        1.0.1 & 2016 & Add more clarification, and correct typos& \cite{lorawanv101} \\
        \hline
        1.0.2 & 2016 & Improve the physical layer section that describes the regional parameters and move it to a separate document & \cite{lorawanv102, lorawanv1022}\\
        \hline
        1.1 & 2017 & Support both passive and active roaming, introduce the join server (JS) for security& \cite{lorawanv11} \\
        \hline
        1.0.3 & 2018 & Import Class B functionalities from v1.1 into the v1.0.x branch, introduce unicast/multicast support to Class B devices. & \cite{lorawanv103} \\
        \hline
        1.0.4 & 2020 & Improving class B functionalities, importing JS functionalities from v1.1 for security enhancement &  \cite{lorawanv104} \\
        \hline
    \end{tabular}
    \label{tab:LoRa_versions}
\end{table}
\section{Interference Management Schemes}\label{sec:interference}
\begin{figure*}[t]
    \centering
    \includegraphics[width=0.98\linewidth]{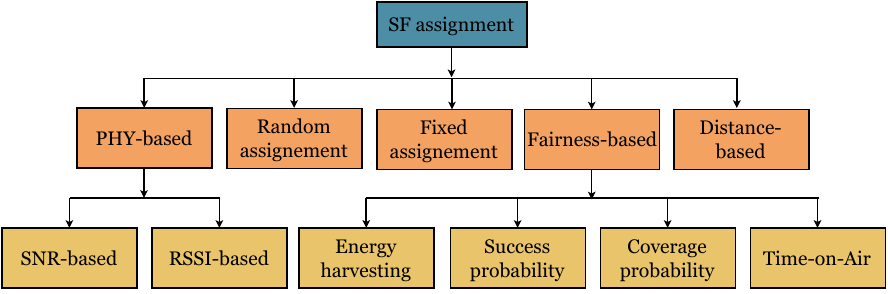}
    \caption{Classification of different SF assignment schemes.}
    \label{fig:sf_assignement}
\end{figure*}
The increasing number of connected IoT devices working within the ISM band makes the LoRa communication system the main focus of applications, where the high interference level is the major challenge for efficient deployment \cite{Moy2020}. In contrast, applications operating in a controlled spectrum incur less interference since a single operator controls and dominates the QoS management task. Thus, future IoT networks designed to work with LoRa should consider the significant impact of different types of LoRa interference and the characteristics of the propagation environment. As such, a framework that estimates reliable network coverage based on real interference measurements was designed in \cite{AlHomssi2021} depicting the spectrotemporal behavior records of the shared band traffic acquired by a software-defined radio.

In LoRa-based IoT networks, the ED may transmit its source data several kilometers to the GW. However, when considering TX in a dense environment, multiple sources interfere with the original signal adding noise and drastically reducing the LoRa-based system performance. Various studies have evaluated this technology for long-distance communication and considered different types of interferences. The immunity region where the interference impact does not reach LoRa communication was identified in  \cite{Marquez2020} through the empirical model of distance-interference trade-off. Their experiments showed that 14~dBm interference level resulted in total packet loss. The LoRa performance in terms of error probability was studied in \cite{Baruffa2020}, where the accurate error probability of the received LoRa chirp was estimated through an additive white Gaussian noise channel. In contrast, some studies \cite{7797659, 8392707, 9014148, 8903531} have provided bit error rate (BER) expression for orthogonal and quasi-orthogonal LoRa signals. To identify the interference level that may influence IoT ED deployment in the cities of Aalborg and Denmark, the authors of \cite{7925650} performed a measurement study using the European ISM band of 863-870~MHz. The outcome of this study identified that signals above 863-870 MHz have 22\% to 33\% probability of interference. 

The authors of \cite{10.1145/2988287.2989163} discussed the limitations of the LoRa ability regarding providing collision-free communication for a great number of devices deployed in a wide area with a low number of GWs. These devices must share the communication medium to transmit their captured data. Therefore, LoRa employs a combination of its features such as SF, BW, CR, and CF. Nevertheless, a limited number of devices can be supported through the standard LoRa modulation, which presents the capacity limits of the LoRa communication system. When surpassing this limit, interference occurs in the system decreasing its performance  \cite{10.5555/3324320.3324402, 10.1145/3293534, 10.1145/3416010.3423228}. In this direction, a detection scheme of the inter-system interference was designed using the density ratio estimation \cite{Aihara2020}. Generally, interference facing LoRa signals were classified into two categories: co-technology interference, where two LoRa-enabled devices transmit data to the same GW via the same communication channel, and inter-technology ISM interference, which depicts the coexistence of different technology besides LoRa communicating via the shared unlicensed ISM bands \cite{8620142, 8422801, 10.5555/3108009.3108093, 7925510, 8115772}. The following section discusses various interference management schemes for the LoRaWAN.

\subsection{SF Assignment}
The SF assignment allows LoRa to control its transmission sensitivity using an ADR scheme that also manages other PHY layer parameters, such as the TP of the ED and the CR  \cite{10.1145/3330482.3330500}. For example, a high SF corresponds to a few chirps sent per second, which means a few data are encoded per second. Therefore, a high SF allows LoRa-enabled EDs to transmit over long-distance but with high collision probability due to the low DR. For a scalable and fair LoRa network, a relay control scheme was suggested in \cite{8430542}. Fairness consists of equalizing the success probability at each SF region while the scalability is allowed by multiple relay nodes using a similar SF to overcome interference in dense star topology-based networks, where EDs communicate directly with the GWs to transmit the source data through the LoRa communication channel. 
The analytical model of the success probability proposed in this work to manage the relay operation considers the different independent factors of the signal-to-interference ratio (SIR) and SNR. Furthermore, the coverage probability through the overall network and the minimum success probability corresponding to each SF region were maximized to ensure fairness. The RSSI values were used in the relay selection mechanism, which helps decide whether to forward the received packet from the source ED.

LoRa performance enhancement was the target of \cite{8115779, Garlisi2021a} by designing an efficient SF allocation scheme based on the RSSI values and the SIR. A heuristic SF allocations method was designed in \cite{8115779} to prove the need to assign larger-than-needed SFs. Sequential waterfilling was used to design a sophisticated SF assignment scheme \cite{Garlisi2021a}, which is an enhancement of the previous work \cite{8115779} by the same authors. Fairness was ensured by equalizing the time-on-air of the transmitted packets using different SFs. In addition, SFs across different GWs were balanced considering the channel capture effect. Stochastic geometry was used to design a heuristic SF allocation algorithm \cite{Saluja2021} where the expression of the packet success probability was derived for co-SF interference scenarios regarding the SNR values. 

Similarly, success and coverage probabilities were estimated in \cite{9244125} while reducing the computation complexity required to obtain these probabilities for any interference type compared to the previous work. The study did not consider the imperfect LoRa feature orthogonality. Fairness concerning the energy harvesting time duration was targeted in \cite{Benkhelifa2021} by optimizing the SF assignment and TP for all LoRa-enabled EDs. The energy harvesting time and allocated power are optimized for single or multiple uplink TX. For the first case, the bisection method was employed to optimize the allocated power, while suboptimal power allocation was obtained using the concave-convex procedure.
\begin{table*}[]
  \centering
  \renewcommand{\arraystretch}{1.6}
  \caption{Summary of SF assignment schemes for scalable LoRaWAN networks}
  \begin{tabular}{p{2em} p{3em} p{10em} p{8em} p{6em} p{5em} p{17em}}
  \toprule
  \textbf{Work} & \textbf{Mobility} & \textbf{Fairness} & \textbf{Evaluation metrics} & \textbf{Assessment methodology} & \textbf{Approach} & \textbf{Main results}\\
  \bottomrule
  \cite{8430542} & \xmark & Equalized the success probability at each SF region & SF-dependent SIR & Monte Carlo simulation & Stochastic geometry \cite{4215725} & Derived SIR distributions through aggregating the co-SF and inter-SF interference power to capture the uplink outage \\
  \hline
  \cite{8115779} & \checkmark & Equalized the ToA of transmitted packets through fair SF allocation & data extraction rate, throughput & LoRaSim Simulation & Ordered waterfilling Technique \cite{6223460} & Outperformed the legacy ADR in terms of bit rate in the case of high traffic loads \\
  \hline
  \cite{Garlisi2021a} & \checkmark & Unfair DER between EDs due to the use of capture effect & Data extraction rate, Network density & LoRaSim Simulation & Sequential waterfilling Technique \cite{6223460} &  Enhancing the legacy ADR in terms of scalability and its robustness against different operating/load conditions\\
  \hline
  \cite{Saluja2021} & \checkmark & Not considered & Packet success probability & Monte-Carlo simulations \& Prototype experiment & Stochastic-geometry \cite{6365639} & Improved the network scalability regarding the size of the packet and cell radius, for real-life applications. \\
  \hline
  \cite{9244125} & \xmark & Not considered & Success and coverage probabilities & Monte Carlo simulation & Stochastic geometry &  Model with a low level of complexity that achieves high performance under perfect orthogonality without considering SNR.\\
  \hline
  \cite{Benkhelifa2021} & \checkmark & Equalized the energy harvesting time duration between EDs & Minimum throughput rate and harvested power & Numerical results in MATLAB & Min-max Optimization & Outperformed the existing SF assignment schemes in terms of minimum data rate. \\
  \hline
  \cite{Kumari2020} & \checkmark & Not considered & The waiting time, Throughput & NS-3 simulation & Stackelberg game & Achieved both Nash and Stackelberg equilibrium, validating the obtained results through the NS-3 simulation \\
  \hline
  \cite{Farhad2020} & \checkmark & Not considered & Packet loss ratio, packet success ratio & NS-3 simulation & Heuristic model & Proactively monitored EDs mobility through dynamically optimizing the SF assignment to reduce retransmissions and enhance packet success ratio. \\
  \hline
  \cite{8766738} & \checkmark & Not considered & ToA, energy consumption & Prototype experiment & Heuristic model & Enhanced the legacy ADR in terms of time-on-air and energy consumption \\
  \hline
  \cite{Bapathu2021} & \xmark & Not considered & Frame error rate & Numerical results & Heuristic model & Ensure a low packet error rate over time-varying channels. \\
  \hline
  \cite{Coutaud2021} & \checkmark & Not considered & Data error rate, ToA & Experimental dataset & - & Optimized the transmission parameters including SF considering LoRa channel characteristics to enhance ToA and packet error rate \\
  \hline
  \cite{Staniec2018} & \checkmark & Not considered & Packet error rate & Prototype experiment & Statistical modeling & Three distinct sensitivity zones in the LoRa configuration space were distinguished based on their immunity to interference and multipath propagation \\
  \hline
  \cite{Lim2018} & \checkmark & The average system packet success probability was maximized & Packet success probability & Monte Carlo simulations & Mathematical optimization & Maximized the packet success probability and the ED connectivity  \\
  \hline
  \cite{Hoeller2020} & - & Not considered & Outage probability & Monte Carlo simulations & Mathematical optimization & Enhanced the energy efficiency and network capacity by reducing interference. \\
  \hline
  \cite{8319183} & \xmark & Not considered & Symbol time, SIR & Prototype experiment & Heuristic model & Ensure a good balance between sensitivity and throughput \\
  \hline
\end{tabular}
  \label{tab:sf_assignement}
\end{table*}

The authors of \cite{Kumari2020} proposed sharing the same SF by different LoRa users for a limited time duration. Therefore, a game theory approach was employed to manage the assignment of the same SF for different ED for different time durations of the communication process to avoid interference between ED transmitting simultaneously via the same SF. The solution to this game consists of finding the equilibrium point where all users maximize their utility (also called the gain) corresponding to the minimum interference level. This kind of game is generally used to model the interaction between different players (LoRa users in this case ) to find the best policy that maximizes the overall gain of our system. The mobility of the ED makes the SF assignment a challenging task due to time-variance in uplink TX properties. The authors of \cite{Farhad2020} suggested a scheme to proactively allocate the SF  for each LoRa user, either static or mobile. For mobile nodes, the SFs are re-scheduled based on the received signal strength from the target GW, which reduces the ToA and enhances the packet success ratio by reducing packet loss and retransmission. Only mobile nodes with a specific pattern were considered in the enhanced-ADR \cite{8766738} to provide a dynamic allocation to minimize the packet loss as well as the TX time for each LoRa transmitter.
\begin{figure}[t]
    \centering
    \includegraphics[width=0.98\linewidth]{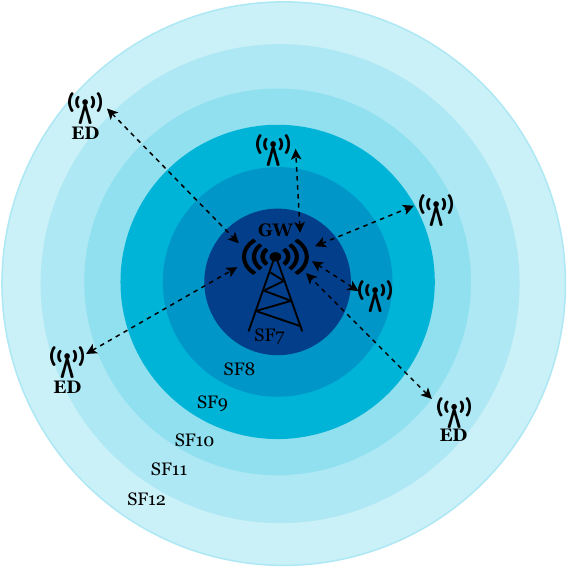}
    \caption{SF assigned to EDs based on the distance from the GW, the nearest ones use the lowest SF to transmit the data}
    \label{fig:SFassignement}
\end{figure}

The authors of \cite{Bapathu2021} answered the question of whether a high SF is needed for robust LoRa communication over a rapid time-varying channel. They benefit from the exponential correlation of Rayleigh fading on the frame error rate of a LoRa receiver employing the CSS modulation to transmit its frame. They observed that the robustness of LoRa ED deploying high SF and transmitting over rapidly-varying channels decreases when the payload size increases. The measurement  in \cite{Coutaud2021} was performed over a public LoRaWAN network deploying multiple GWs to receive data from different LoRa ED deployed in a medium-sized city. The aim was to characterize the communication channel by identifying multipath fading, loss burstiness, frame length, and the FEC required for robust LoRa communication. This method was employed to optimize the performance of the target LoRaWAN network in terms of reliability and time-on-air by adjusting the SF and the number of frame repetitions. Three different sensitivity regions were designed in \cite{Staniec2018} to characterize the immunity of the LoRa communication system against interference and multipath fading issues. This categorization was based on the major LoRa features: SF, BW, and CR. The white region depicts the immunity against both interference and multipath fading; the light-gray region means that the system is immune to multipath issues but vulnerable to interference. The LoRa communication system is susceptible to both phenomena in the dark-gray region. For a summary of all discussed solutions in this section please refer to Table \ref{tab:sf_assignement}.  

Legacy schemes assign the SF according to the distance of the ED from the GW for dense LoRa-based IoT network deployments (as depicted in Fig.~\ref{fig:SFassignement}), such as the equal-interval-based method \cite{7803607}, equal-area-based scheme \cite{Lim2018} and the Random scheme where the SFs are assigned randomly without considering the LoRa ED location. 
When using the equal-interval-based scheme, the interval $[d_s, d_{s+1}]$. The following formula can determine the interval using the equal-area-based scheme $(d_{s+1} * d_s) = R_c/S$, $d_s = R_c (s -1)/S$, where $R_c$ is the region index coverage radius for a single GW and S. Numerous studies have focused on analyzing LoRa performance under different scenarios including massive IoT connectivity, harsh communication environments, multiple and single GWs deployments \cite{Hoeller2020, 8319183, Guo2021}. Contrary to the distance-based SF assignment, the authors of \cite{Hoeller2020} analyzed the performance of ADR, which allocates the TP and SF based on the channel state. Working on single-cell LoRaWAN systems restricts this study to  LoRaWAN deployment scenarios in small towns or villages \cite{10.1145/3416010.3423228}. 

Thus, an analytical LoRaWAN model was described and validated through Monte Carlo simulations. Co-technology interference was covered in the performance evaluation of LoRa receiver \cite{8319183} regarding the advantage of multi-hop LoRa network for ensuring wide network coverage and enhancing indoor penetration compared to the single-hop LoRa network  \cite{10.1145/3479239.3485718}. This advance is due to the use of relays to advance packets into the destination, extending the network capacity using different SFs to transmit to the GW or forward the received packets. In addition, the SIR required to enable SFs orthogonality was evaluated, and LoRa immunity against interference resulted from transmitted time-synchronized packets by multiple ED using similar SF. Experiments were conducted \cite{Guo2021} to evaluate the LoRa packet reception performance. A trade-off between the physical parameter and packet reception performance was identified under a negative SNR. The packet length greatly influenced the LoRa performance; this fact was used to design a TX scheme that manages the reliability, TX delay, and energy consumption. 

\subsection{Interference Cancellation}
Based on the features enabled by the LoRa CSS modulation (SF, TP, CR,  and BW), a LoRaWAN GW is supposed to decode multiple superposed received LoRa signals transmitted using different SFs. Although some studies have revealed that LoRa signals sent with different SFs are not entirely orthogonal \cite{8319183, 10.1007/978-3-319-67639-5_13}, this is true as long as the network density is lower than the LoRaWAN capacity defined by the combination of different PHY layer parameters where LoRa can ensure reliable connectivity through the network. One of the pioneering works to suggest using interference cancellation is \cite{Halperin2007}, where the aim was to enable efficient communication in the presence of concurrent TX and enhance the MAC layer capabilities. The authors of \cite{Halperin2007} explained how realistic the interference cancellation (IC) is and its feasibility in practice, although signal processing schemes should be adapted to follow these changes. A mathematical model presented in \cite{8292748} as an extension of \cite{7974300} models the LoRaWAN channel access process considering the capture effect. The proposed model enhances the network capacity and provides a reliable LoRa-based TX due to its accurate data TX process description regarding the power difference of signals received simultaneously from multiple LoRa transmitters. This model has usually been used in the literature to evaluate the LoRaWAN network performance under different use cases, including the case where they consider the Okumura-Hata propagation model \cite{8372906, 8407095, 8615313, 8633860}.

\begin{table*}[t]
  \centering
  \renewcommand{\arraystretch}{1.8}
  \caption{Summary of interference cancellation schemes against LoRa scalability issues}
  \begin{tabular}{l >{\centering}p{10em} >{\centering}p{5em} >{\centering}p{5em} c c >{\centering}p{8em} c}
  \toprule
  \textbf{Work} & \textbf{Successive interference cancellation} &\textbf{Frame preamble} & \textbf{Propagation loss} & \textbf{Capture effect} & \textbf{Fourier transform} & \textbf{Realistic implementation} & \textbf{Scenario}\\
  \bottomrule
  \cite{Halperin2007} & \checkmark &  & \checkmark &  &  & \checkmark & Downlink\\
  \hline
  \cite{8292748} &  &  & \checkmark & \checkmark &  & & Downlink \\
  \hline
  \cite{8638253} & \checkmark &  &  &  &  &  & Down / Uplink\\
  \hline
  \cite{8385552} & \checkmark &  &  & \checkmark &  &   & Uplink\\
  \hline
  \cite{8904258} &  & \checkmark &  &  & \checkmark & \checkmark & Downlink\\
  \hline
  \cite{9217060} & \checkmark & \checkmark &  &  & \checkmark & & Uplink \\
  \hline
  \cite{Tesfay2020} &  & \checkmark & \checkmark &  & \checkmark &  & Downlink\\
  \hline
  \cite{9148783} & \checkmark & \checkmark &  &  & \checkmark & & Uplink \\
  \hline
  \cite{9282261} & \checkmark & \checkmark &  &  & \checkmark &  & Downlink\\
  \hline
  \cite{Garlisi2021} & \checkmark & \checkmark & \checkmark & \checkmark & \checkmark & \checkmark & Downlink\\
  \hline
  \cite{9644605} & \checkmark &  & \checkmark &  & \checkmark & \checkmark & Uplink\\
  \hline
  \cite{9155433} &  & \checkmark &  &  & \checkmark & \checkmark & Uplink\\
  \hline
  \cite{Xia2020} &  & \checkmark &  &  & \checkmark &  \checkmark & Downlink\\
  \hline
  \cite{DeSouzaSantAna2020} & \checkmark &  & \checkmark &  &  &  & Downlink\\
  \hline
\end{tabular}
  \label{tab:Interference-Cancellations}
\end{table*}

The authors of \cite{8638253} suggested two methods to decode the overlapped LoRa signals. The first works on desynchronized transmitters, and the second requires the transmitters to be slightly synchronized. Their performance evaluation proved that the proposed algorithms enhance the overall throughput, especially the first algorithm, which improves throughput significantly, whereas the second one only decodes the strongest signals. The first algorithm regarding the desynchronized transmitters consists of the following steps to decode the received overlapped frame: the receiver uses preamble detection to detect the reception of two overlapped signals and identifies the symbol frontier of each one. The second step compares the updated previous and new frequencies based on the data decoding. This comparison helps the receiver identify the transmitters and where their symbols start. In the same direction, \cite{8888038} designed a mLoRa protocol implementing successive interference cancellation (SIC) to decode the received frames successfully. Through experiments performed using the USRP testbed, the proposed scheme showed his capability to decode three concurrent frames received simultaneously. Likewise, \cite{9060885} presented a new LoRa receiver equipped with a commercialized LoRa chip to deal with colliding LoRa signals.

The capture effect and SIC explored in \cite{8385552} to build a robust LoRa-based system against collision to support channel sensing unable to solve this issue due to the wide coverage area of the GW. In this context, the SIC uses the sum of received signals to identify simultaneous TX successfully. At the same time, the capture effect allows for successfully demodulating at least one signal in the presence of multiple colliding signals. Both methods improve the overall throughput of LoRa-based systems, although good results were obtained when the power difference of the received signals exceeded a certain threshold. Moreover, the LoRa-based system capacity limit was studied by analyzing the results of the packet reception rate as a function of the network size. A commercialized LoRa chip was employed to process collision, making the proposed solution \cite{8904258} more realistic and suitable for implementation in an already existing network to enhance spectral efficiency. The enhanced LoRa-Like receivers estimate the time shift between them from two received signals to enable processing unsynchronized collided signals using fast Fourier transform (FFT) representations. The simulation results and the mathematical expressions provided in this work showed that the frame preamble could be used to identify the value of desynchronization between the two received LoRa signals.

In the same direction, multiple studies have focused on enhancing LoRa-like receiver \cite{9217060, Tesfay2020, 9148783} to successfully decode LoRa signals received simultaneously and using similar SFs. In \cite{9217060} the legacy LoRa network capacity increased by 20 to support more EDs through designing a complete receiver, including an interference cancellation scheme, channel estimation, and packet detection models. Implementation issues arise as a major drawback of the proposed LoRa receiver. Another uplink scheme was designed in \cite{Tesfay2020} using the SIC, and the particular structure of the received superposed LoRa-like signals. The proposed schemes were implemented in the GWs due to their unlimited energy consumption. Thus, the energy consumption of EDs was reduced, and their spectral efficiency was enhanced due to the low number of re-transmitted packets. In contrast, \cite{9148783} designed a new multiuser detector to improve the downlink in LoRa-like networks. This is due to the downlink's capability to limit the number of acknowledgments sent after successfully receiving packets, which directly influences the reliability of the LoRa-based network. The designed scheme was inspired by the non-orthogonal multiple access (NOMA) \cite{7676258} that reduces the impact of LoRa network sequential TX, the duty cycle, and the half-duplex mode. Using the NOMA technique to build the proposed scheme overcomes the interference cancellation issue, namely, the residual error reduces its complexity. Furthermore, NOMA was employed in \cite{9282261} to enhance the downlink for CSS communication combined with a SIC algorithm in the presence of concurrent CSS signals received simultaneously. The desynchronization time was managed to assign different powers to the transmitted packets.

LoRa receiver with syNchronization and cancellation (LoRaSyNc) was designed in \cite{Garlisi2021} to enhance LoRa network capacity limit by extending the traditional demodulation procedures by synchronizing superposed signals and adding a clock tracking scheme to the frame reception procedure. The authors of \cite{9644605} addressed the stability of LoRaWAN  by designing a SIC LoRa receiver. This two-user detector uses the bit-interleaved coded modulation to detect and cancel the strongest interfering signal, considering a soft-demodulator and soft-decoder to attain acceptable error rates regarding the low SNR regime of LoRa communication. The online concurrent TX scheme for LoRa GW \cite{9155433} consists of three steps: identifying the frame preamble, detecting the start-of-frame-delimiter, and packet decoding using only the legacy LoRa’s (de)modulation. FTrack presented in \cite{Xia2020},  \cite{10.1145/3356250.3360024} adopted a technique to distinguish between colliding LoRa signals using their preamble and chirp characteristics. The model built in \cite{DeSouzaSantAna2020} resolves collided packets jointly by considering the SIC and a stochastic geometry model of LoRaWAN. Table \ref{tab:Interference-Cancellations} summarizes the interference cancellation schemes proposed to alleviate LoRa signals interference using different approaches.

\subsection{GW Densification}
GW densification arises as an effective method to face the increased IoT ED density in the target area due to the growing interest in IoT devices, especially in the downtown city's popular and crowded urban environment. Thus, LoRaWAN networks have resorted to dense GW deployment, which should be performed while avoiding the considerable interference between the massive IoT ED and densely deployed GWs. New densification strategies should be designed since the existing ones suggested for cellular networks are unsuitable for the LoRaWAN use case. This is regarding the advanced coordination within and between cells to manage the radio resources, which is not the case for LoRaWAN \cite{Raza2017}. 

A GW densification approach was suggested in \cite{9149081} by analyzing multi-cell LoRa networks to scale them. To this end, multiple GWs were deployed for free co-SF interference in dense networks. The proposed model considers the interaction between GWs and some specific properties of the LoRa technology. The coverage and scalability of the deployed LoRa GWs were analyzed mathematically, such as using the homogeneous Poisson point processes to  distribute GWs and EDs uniformly and randomly in the target area. The network performance was evaluated based on the coverage probability derived from the stochastic geometry tools \cite{haenggi2012stochastic} to unveil the positive impact of GW densification on the network coverage as well as its scalability.
\begin{table*}[!t]
  \centering
  \renewcommand{\arraystretch}{1.8}
  \caption{Summary of GWs densification proposals suggested to cope with the increasing IoT network density}
  \begin{tabular}{p{4em} p{6em} p{5em} p{6em} p{5em} p{6em} p{22em}}
  \toprule
  \textbf{Proposal} & \textbf{Scenario} & \textbf{Traffic type} & \textbf{GWs distribution} & \textbf{Analysis type} & \textbf{Performance Metrics} & \textbf{Description}\\
  \bottomrule
  \cite{9149081} & Industrial IoT & Uplink & PPP & Mathematical analysis & Coverage, Success probabilities & Model the interaction between multiple GWs as well as some interesting Semtech specifications to enhance the network coverage and support scalable LoRa network \\
  \hline
  \cite{Yu2020} & massive-IoT & Uplink & Optimized & Monte Carlo simulations & Coverage probability & Optimized the GW position to increase the network coverage and enhance network performance while considering all types of interference in LoRa networks.  \\
  \hline
  \cite{Chevillon2021} & Urban & Uplink & PPP & Theoretical, numerical analysis & Coverage, success probabilities; SIR & Analysed the impact of neighboring networks on the performance of LoRa networks through modeling the interfering networks using $\alpha$-stable distributions, and considering the coverage probability of different GW densities. \\
  \hline
  \cite{9253425} & Dense network & Uplink, downlink & 
UMED distribution & Mathematical analysis & Time on air, Data error rate & Developed LoRa communication channel model from a real-life experiment to estimate DER in a network of multiple GWs, and ensure a high level of reliability over the traditional ADR. \\
  \hline
  \cite{s21196488} & Smart city & Uplink, downlink & Optimized & FUOTAsim simulator & Update, Energy efficiency & Improved the time and energy efficiency of firmware updates in dense LoRaWANs \\
  \hline
  \cite{Sun2021} & Smart city & Uplink, Downlink & Optimized & Numerical results & Throughput, GWs density, Energy efficiency & GWs planning scheme was proposed in order to increase the energy efficiency and throughput of hybrid LoRaWAN networks \\
  \hline
  \cite{Magrin2021} & Industrial IoT & Uplink & Uniform & Numerical results & Confirmed Packet Success Rate & Achieved a packet success rate of over 90\% with limited communication delays in multi GWs deployment through considering both confirmed and unconfirmed traffic using different EDs classes  \\
  \hline
  \cite{bouazizi2020spatiotemporal} & Dense network & Uplink & PPP & Monte Carlo simulations & Transmission success probability & A spatiotemporal model of multi-GW deployment was developed considering imperfect SF orthogonality to increase transmission success probability\\
  \hline
  \cite{Aftab2020} & IoT network & Uplink & Deterministic & Numerical results & Coverage probability & Analyzed LoRaWAN scalability in multi-GWs deployment in the presence of co-SF interference using stochastic geometry \\
  \hline
  \cite{van2017scalability} & Massive IoT & Uplink, downlink & Deterministic & NS-3 simulation & Packet delivery ratio & NS-3 simulations have demonstrated that increasing the GW density cannot eliminate all issues related to scalability since duty cycle restrictions remain applied for GWs, restricting downstream traffic. \\
  \hline
\end{tabular}
  \label{tab:gw_densification}
\end{table*}

The coverage probability of edge EDs using one node distributed randomly through the network was deeply studied in \cite{Nguyen2020} under different fading models. A closed-form expression coverage probability was calculated when considering the Rayleigh fading channel, while only an approximation was computed for the Nakagami-m fading channel. This study concluded by suggesting the intelligent reflecting surface and an alternative to the legacy relay strategy for LoRa networks \cite{8888223}. The coverage of LoRa links under the presence of multiple GWs was studied in \cite{Chevillon2021}, considering the interference from the LoRa network itself, called co-SF interference. The coverage probability of EDs under these interferences was expressed based on the stochastic geometry considering the interference from underlying technologies (such as cellular communications) as impulsive noise \cite{8761783} and was modeled as an $\alpha$-stable distribution. The study concluded that the existence of underlying technology in the neighborhood of the LoRa network drastically reduces its coverage probability, and this impact increases when considering space GWs deployment. An ADR mechanism was suggested in \cite{9253425} to manage the benefit of deploying multiple GWs and using FEC to enhance the reliability of the LoRaWAN network concerning its load. Thus, FEC is defined as the frame erasure rate, which is the physical loss between an ED and a single GW without considering the frame repetition, showing a significant influence on defining the optimal operating point of LoRaWAN networks. The proposed scheme consists of adjusting the TX parameters of the ED communicating through a channel modeled using experimental measurement over a public LoRaWAN network in the presence of multiple GWs and FEC. The data loss ratio between the EDs and the application server depicted by the data error rate was derived from the proposed channel model to evaluate the performance of the enhanced ADR scheme.

Network densification was suggested in \cite{s21196488} to support firmware updates in dense LoRaWAN networks. Knowing that the firmware update is necessary for EDs to accomplish their tasks efficiently allows upgrading the ED firmware, enhancing their functionalities, and adding more security features. Although the firmware update expanded the network lifetime and improved its performance, it's considered a challenging task due to the huge number of EDs in remote areas requiring periodic software updates. A new solution was designed in \cite{alliance2019fuota} for this purpose: firmware update over the air (FUOTA), which allows performing firmware updates remotely via a wireless medium. Planning and finding the minimum number of GWs and their optimal location to increase the number of EDs covered by the network will be essential for LoRa deployment in massive IoT. Due to the impact of GWs on defining LoRa performance, the authors of \cite{Sun2021} designed a GW planning scheme for hybrid LoRa networks.
Moreover, the GW coverage determines the SNR values, DR of EDs under control, and overall throughput. In hybrid LoRa networks \cite{8875418}, GWs work on opposite frequencies in the downlink and uplink to cover the EDs working with the same and different frequencies. A hybrid LoRa network suggested overcoming the self-interference appearing in the deployment of full-duplex commercial GWs \cite{7956006}.

Multi-GWs deployment was considered in \cite{Magrin2021} to evaluate the performance of the LoRaWAN in the industrial IoT (IIoT), where different classes of EDs were used to ensure a low-cost and efficient industrial process. The analyses were performed using realistic measurements based on \cite{tanghe2008industrial} covering both confirmed and unconfirmed traffic and a non-standard channel plan. Through this study, LoRaWAN showed its power to perform 90\% of packet success rate for IIoT sensing applications. The authors of \cite{bouazizi2020spatiotemporal} modeled multi-GWs LoRa networks using the Matern cluster process to evaluate its uplink TX based on a stochastic-geometry framework. The Laplace transformation of the intra- and inter-cluster interference was analyzed regarding the multi-cell topology imperfect orthogonality between different SFs. Similarly, authors in \cite{Aftab2020} discussed the LoRaWAN scalability by using the stochastic geometry framework to analyze the impact of deployment of multiple LoRa GWs on increasing the number of covered ED. The NS-3 simulation in \cite{van2017scalability} revealed that increasing the GW density can not eliminate all the scalability issues since the duty-cycle restriction remains applied for GW which restricts its downstream traffic. Please refer to Table \ref{tab:gw_densification} for a summary of all GWs densification solutions above.

\subsection{Key Insights.}
One of the first solutions proposed to deal with LoRa co-technology interference arising due to high connectivity demands is to enhance SF assignment schemes. The standard LoRa ADR functionality deployed to manage DR by tuning LoRa physical parameters has reached its maximum capacity. Therefore, new approaches have been proposed to take over the distance-based scheme \cite{7803607, Lim2018} that assigns the SF to LoRa EDs based on their distance from the GW. This method suffers in low-dense networks with a non-uniform distribution of EDs where a high SF is allocated to enable long-range communication, draining the energy of EDs. Fairness has been considered in \cite{8430542, 8115779, Benkhelifa2021} in terms of success probability, time-on-air, and energy harvesting, respectively, to address this deficiency. However, these fair schemes can not ensure a reliable LoRaWAN with an efficient data communication system because the time-variant event occurrence rate, data arrival rate, or mobility of EDs were not considered. Complex schemes were proposed to manage the SF assignment over a multi-cell LoRaWAN \cite{Hoeller2020, 8319183, Guo2021}. Cellular networks should inspire solutions in this context, and game theory can be used to model the interaction between GWs. 

Interference cancellation has demonstrated its ability to deal with the non-orthogonality of SFs and mitigate co/inter- SF interference \cite{8319183, 10.1007/978-3-319-67639-5_13} but it requires adaptable signal processing schemes \cite{Halperin2007}. The capture effect where only the strongest signals are decoded was used to assist the SIC that allows recovering packets with weak signals \cite{8292748, 7974300, 8385552}.
Moreover, the preamble and chirp are some characteristics added to the original signals for synchronization \cite{Garlisi2021} and to distinguish between similar received signals \cite{Xia2020, DeSouzaSantAna2020}. The LoRa signals presentations were suggested \cite{8904258} to enhance the efficiency of LoRa-like signal detection by estimating the time shift of the signals and canceling the strongest interfering signals. The LoRa-like receivers capable of decoding multiple LoRa-like signals received simultaneously have been suggested to enhance the LoRaWAN capacity \cite{9060885, 9217060, Tesfay2020, 9148783}. Although the promising results obtained by adding more features to LoRa transceivers increase the number of simultaneous LoRa-like signals that can be successfully decoded, this increase results in complex and high-cost EDs, which are not tolerated in LPWAN technology.

Instead of being restricted to the star topology where only one GW is deployed to relay collected data from LoRa ED to the NS, a multi-cell-based LoRaWAN was suggested to enhance the scalability of the network and allow efficient dense LoRaWAN deployments \cite{Raza2017, 9149081}. This is also known as GWs densification, which has been studied considering the LoRa communication channel models \cite{Nguyen2020, 8888223} while designing realistic coverage probability derived from stochastic geometry tools \cite{haenggi2012stochastic}. An efficient GW densification should be meticulously designed to avoid introducing new interference types resulting from miss-coordination between the different GWs or other  technologies \cite{Chevillon2021, s21196488}.  

\section{Collision Avoidance Mechanisms}\label{sec:Collision}
This section discusses various solutions in the literature to avoid collision between concurrent TX. The aim is to enhance the LoRaWAN capacity to cover more EDs and fit the 
 massive IoT requirements in terms of connectivity demands. Thus, some studies have leveraged LoRa features to design various logical and frequency channel allocation schemes. In contrast, have others focused on scheduling LoRa TX. In addition, numerous works were inspired by the traditional slotted MAC and carrier-sensing medium access methods to design new features suitable for LoRaWAN specifications. 
  The proposed solutions in this section were inspired by conventional computer networks with some adaptations to meet the LoRaWAN requirements in terms of energy consumption, duty-cycle restriction, and a low DR. The solutions consider the feature offered by LoRa communication to manage the DR (using the standard ADR mechanism). The following sections discuss each of these schemes separately.
\subsection{Windowing \& Channel Allocation}
The severe data collision occurring during the data collection process from LoRa EDs can be alleviated using multiple communication channels. Thus, devices with different traffic loads may be assigned to different channels to tackle the propagation time-variance issue. An allocation scheme of the channel and back-off window size was designed in \cite{Shen2021a} considering homogeneous and heterogeneous LoRa networks. The objective is to enhance the overall throughput regarding the end-to-end latency for the two scenarios above. In the homogeneous LoRa network, a single channel is considered with the same traffic load, which means an identical arrival DR, whereas, in the heterogeneous one, multiple channels are considered with various traffic loads. The formula to derive the LoRa network capacity corresponding to the optimal packet TX probability was defined and validated through experiments and simulations. The scheme showed high capability for assisting massive IoT networks by ensuring a high throughput and low end-to-end latency compared to the legacy LoRa network and existing enhancement.  

LoRa offers users six SFs to transmit data, which means a bounded number of virtual channels are available for a specific time duration. This limited number of virtual channels restricts the number of simultaneous uplink TX to the GWs and defines the network capacity such that a maximal number of devices can transmit correctly. Packet collisions frequently occur when exceeding the network capacity, resulting in a drastic decrease in network performance. The game theory was proposed \cite{9149006} to model the trade-off between the network capacity and virtual channel allocation. The aim was to restrict the use of each SF for a limited time duration while minimizing the waiting time for EDs. The LoRa user requirement was considered in \cite{9205428}, and such multiple channels were scheduled to avoid collision while satisfying the constraints of duty-cycle co-SF interference and the user requirements. The authors assumed perfect SF orthogonality to achieve efficient joint channels and SF allocation that extend the communication range and significantly reduce packet collision.  

The LPWAN adopts asynchronous random-access protocols, such as the pure Aloha protocol in the MAC layer. These protocols are known for their high packet collision probability, where multiple nodes may send their collected data simultaneously via a similar frequency channel. Carrier sensing and centralized resource management were suggested to deal with the packet collision issue. Although the large coverage area of LoRa networks makes carrier sensing infeasible, the high energy consumption of centralized schemes makes this solution unsuitable for LoRa networks. The authors of  \cite{Kaburaki2021} suggested using machine learning to automatically manage the TX timing and avoid unnecessary, redundant packet TX. The proposed scheme based on Q-learning does not require additional signals to effectively manage the TX probability and timing to avoid packet collision. In the first step, nodes detected a new event using Q-learning to determine its TX timing. Multiple nodes may perform this step when simultaneously detecting the same event. Packet collision is avoided in this step by allowing TX time shifts of the same event. The additional step defines the event packet TX as a probabilistic process to enable the correlation of the same event TX, which reduces the uplink load and helps avoid packet collision.

Besides packet collision, window mismatching is a severe issue facing LoRa concurrent TX resulting in high ACK retransmission by GWs to acknowledge the received packets, draining the residual energy of the EDs. This issue does not appear when a single ED tries to transmit its data to the GWs and successfully acknowledges the reception by sending an ACK packet during its TX windows that match the RX windows of the sender. However, when multiple EDs concurrently send their data to the GW, in this case, the GW can only acknowledge the received data when the TX through the uplink is completed. This ACK sent by the GW misses the RX windows of the EDs because the senders switch immediately from TX to RX windows after sending the collected data and wait for an acknowledgment from the GW. After a while, the sender closes its RX windows to save energy and retransmit its data. From this scenario, we can deduce that the throughput does not objectively evaluate the LoRa network. For this, goodput should be considered for LoRa network evaluation since the throughput does not necessarily include only the goodput; duplicated packets may be transmitted, resulting in BW waste and useless energy consumption. An optimization solution designed in \cite{Xu2021} to efficiently manage the RX windows to improve the goodput of LoRa concurrent TX while considering the reception probability of downlink TXs.

In \cite{Chinchilla-Romero2021}, the authors designed resource assignment schemes for different LoRa use cases, such as smart homes, smart healthcare, smart metering, and smart agriculture to improve the network capacity. The schemes leverage the SF orthogonality and multi-channel structure. The space defined by channels and SFs is divided into multiple resource blocks used by EDs to transmit their data simultaneously without collision. Distributed scheduling was used for resource assignment considering a realistic propagation model covering both indoor and outdoor environments (rural and urban). Logical channels modeled in \cite{9016433} using both BW and SF to reflect the SFs orthogonality and the bit rate difference. Each BW and SF pair depict a logical channel with a different $R_b$ value due to the chirpiness.

Furthermore, a logical channel assignment scheme inspired by the minimum weighted sum bin packing problem was suggested to reduce energy consumption and response time for LoRa-based massive IoT networks. The authors of \cite{Yu2020} described an adaptive LoRa sub-band allocation policy, which consists of assigning different adjacent channels of 125~kHz to the EDs located at the edge of SF regions—knowing that LoRa BW of 1 MHz can be split into eight sub-bands of 125~kHz considered in \cite{Yu2020} as adjacent channels. The evaluation of this policy was performed statistically in terms of error, rate distribution, and power consumption for a LoRa network where EDs and GWs are deployed in the target area following a random Poisson point process \cite{baccelli2009stochastic, andrews2016primer}.

\begin{figure*} 
    \centering
  \subfloat[\label{fig:cluster}]{%
       \includegraphics[width=0.48\linewidth]{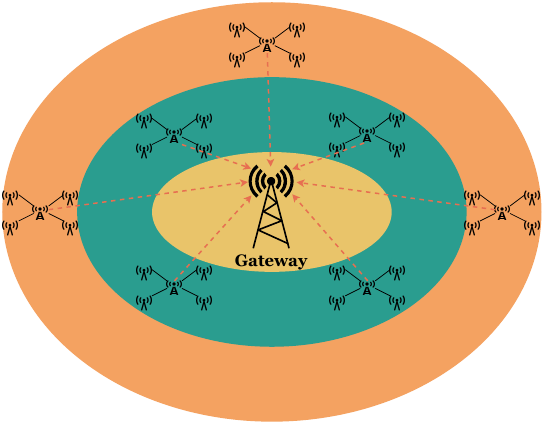}}
       \qquad
  \subfloat[\label{fig:multihop}]{%
        \includegraphics[width=0.48\linewidth]{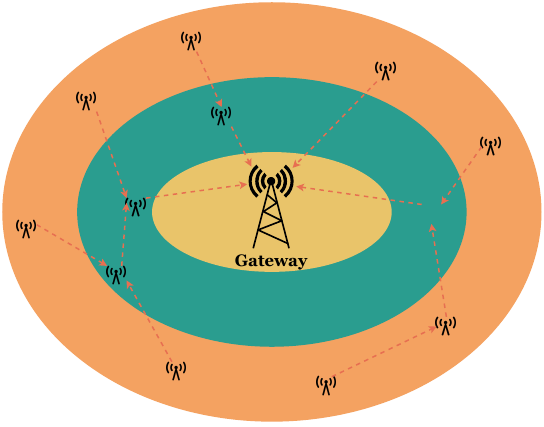}}
  \caption{Beyond star topology: (a) EDs grouped into cluster only the cluster head communicates the collected data to the GW via LoRa communication link (b) EDs use multi-hop communication (instead of peer-to-peer communication adopted in the legacy star topology) to extend their transmission range and transmit collected data to the GWs }
  \label{fig:beyond} 
\end{figure*}

\subsection{Beyond Star Topology}
Traditional LoRa networks adopt start topology to connect EDs deployed around the GWs, restricting its coverage area and limiting its capacity for covering more EDs. Multihop communication can be considered based on the LoRa features to manage direct device-to-device links to cope with this issue. For example, the SFs not only provide flexibility regarding the DR and sensitivity but also give another dimension to the multi-access channel. Using this new dimension, the authors of \cite{8637935} focused on improving the LoRa network capacity by adopting multi-hop communication where the data traffic is off-loaded into multiple subnetworks. In this scheme, paralleled TX is realized by clustering EDs based on the SFs, e.g., each subnet uses separate SFs to communicate its data (Fig.~\ref{fig:cluster}). The clustering is performed taking into account the connectivity between subnets, maintaining a balanced traffic load due to the SF asymmetry, and  minimizing the hop count to decrease the TX airtime. The Bottom-up Breadth-First-Search algorithm was used to extract the farthest nodes from the root, and top-down breadth-first-search was applied to insert them into a new sub-tree \cite{8253754}.   

By keeping the baseline star topology of LoRa, clustering was considered in \cite{AsadUllah2019} to build a reliable LoRa network with low energy consumption. For example, the k-means clustering scheme was proposed for SF allocation in a large-scale LoRa network using a single GW. Fairness was ensured in performance and SF usage, such as minimal difference occurred in the performance of the worst- and best-case nodes, fairly distributing the energy consumption between nodes, and increasing the network lifetime. 

In contrast, SFs are equally used throughout the network. Unlike the existing study where the SF areas were defined based on a constant step distance from the GW \cite{7803607, 8372906}, different numbers of clusters with a maximum range were adopted for distinction. In \cite{Valtorta2019} LoRaWAN packets were collected and processed for profiling and grouping IoT EDs to improve LoRa-based technology operation and ameliorate network anomaly detection. Specifically, the K-means algorithm was used for clustering IoT EDs based on their behaviors and the operations performed over the networks. Machine learning tools were deployed to provide accurate profiling based on the collected data from EDs connected to the LoRa network. 

Based on the LoRa features, LoRaWAN coverage can be expanded at the expense of low throughput and increased energy consumption. Multi-hop communication was used \cite{Farooq2020} to ensure simultaneously high throughput and large coverage (Fig.~\ref{fig:multihop}). The LoRa modulation capabilities define the layers radius, where the network is partitioned into multiple layers, each comprising multiple operational GWs. The traditional LoRa network does not support layering, thus multiple messages should be defined to be exchanged between the GWs and EDs to incorporate the proposed functionalities. LoRa TX parameters were clustered in \cite{9297284} based on the QoS performance in terms of the bit error rate, ToA, and RSSI to deal with the heterogeneity of the IoT application requirements. As such, a set of TX settings providing a similar QoS were grouped into the same cluster to adapt the wireless communication link regarding the IoT application requirements concerning the QoS metrics, as mentioned. Thus, a fuzzy C-means clustering scheme was used to cluster random transmission settings. The priority scheduling technique \cite{Alenezi2020} was proposed for implementation in LoRa GWs to group EDs based on the TX priority. This scheme was built based on unsupervised learning algorithms that allowed the GW to schedule the TX of EDs regarding the corresponding cluster. The aim was to scale up the network density by avoiding collision, consequently reducing the TX delay and energy consumption resulting from continuous retransmission attempts.

Fairness in terms of the throughput was sought in the clustering scheme in \cite{Muthanna2021}, where the nodes were grouped based on the maximum served traffic using two common classification methods, namely K-means, and FOREL. In addition, the throughput distribution in the clusters was modeled to provide an accurate estimation of the real throughput capacity in each cluster. That the EDs communicate their data to all GWs in their range, increasing useless energy consumption resulting from the redundant TX. The impact of the cluster range, EDs TX range, as well as network density were studied in \cite{Gupta2022}. Clustering was proposed to avoid redundant TX in dense networks to reduce energy consumption and increase the network lifetime. The proposed method seeks the shortest multipath route that connects the EDs to the GW and the cluster head. LoRa multi-hop network suggested previously in \cite{Liao2017} to enable concurrent TX and build a robust and efficient LoRa network considering the LoRa PHY layer standard. Network slicing was combined with software-defined networking \cite{8806711} for an enhanced network configuration for large-scale scenarios, where LoRa parameters are tuned to provide an optimized QoS threshold for each virtual network slice. The LoRaWAN adopting slicing architecture displayed promising results through the NS-3 simulation in terms of reliability for the QoS thresholds in dense networks. Hybrid deployment was also used to enhance the LoRaWAN network capacity \cite{Jiang2021, Edward2021}. In \cite{Jiang2021}, ANT technology \cite{6616827} was enabled with the LoRa to enable short-range and ultra-low power communication for data gathering in ultra-dense LoRa networks. An interleaved chirp spreading LoRa-based modulation was introduced in \cite{Edward2021} to work in parallel with the standard LoRa modulation.  

\subsection{Carrier Sensing}
The ease of deployment of LoRa networks as specified in \cite{sornin2015lorawan} aims to ensure the coexistence of multiple networks as well as thousands of EDs based on the features provided by the LoRa PHY layer, such as multiple logical channels and the SF orthogonality. However, regarding the set of features, LoRaWAN remains in the category of Aloha protocols, which are well known for their limitations as indicated in \cite{1096124}. Moreover, long-range communications are known for their low throughput (about 100~bps to 30~kbps) and long time-on-air TX of transmitted messages. Thus, they play a crucial role in increasing the collision probability between concurrent TX in addition to their limitation concerning the duty cycle regulation imposed on LoRa communication. Low-cost LoRa testbed experiments were performed in \cite{8376997} to analyze the impact of using the carrier sensing mechanism on reducing collisions between simultaneous TX. The comparison between the real-world testbed measurement and simulation results obtained by the NS-3 simulator revealed that using CSMA to enhance LoRaWANs in terms of collision avoidance succeeded in increasing energy consumption \cite{8422800}. However, in dense network deployments, the legacy LoRa network presents significant energy consumption compared to CSMA whereas the latter increases the overall throughput taking advantage of the limited European Telecommunications Standards Institute duty cycle restriction \cite{9139047}.

The authors of \cite{8766700} investigated the use of a listen-before-talk mechanism to enhance the LoRaWAN channel utilization and efficiency due to its ability to determine whether the channel is idle or used. That reduces the collision between simultaneous TX using the same SF and transmitting via the same channel. Modified versions were designed for this purpose, such as the binary exponential back-off (BEB), binary exponential delay, and binary exponential hybrid, which combine the two previous algorithms. These algorithms were used to model the interaction between the channel activity detection scheme and the data TX. The numerical results showed that BEB algorithm has the minimum energy consumption during the channel activity detection process and significantly decreases the collision rate. An advanced CSMA/CA variant was suggested \cite{8966182, 10.1145/3372224.3419200} to ameliorate the LoRa IoT network experience in the context of smart city applications. The variant includes some specifications of this kind of application, and a percentage of hidden nodes was considered, demonstrating promising results regarding various sensing ranges. This approach can help with LoRa-based IoT deployment by anticipating the CSMA performance under the considered environment conditions.

Practical viability of carrier sensing was provided in \cite{9142799} for a LoRa-based wildlife monitoring network. By evaluating the proposed channel activity detection scheme considering both laboratory and field conditions, the authors observed that the payload symbols and preamble remain detectable up to 4~km  from GW. Moreover, performing the channel activity detection eight times during payload frame time can ensure a clear channel assessment, reducing the collision rate. A hybrid MAC protocol based on slot scheduling and multiple listen-before-talk (mLBT) was designed \cite{Hoang2020} to deal with data loss resulting from collision and signal suppression. The aim was to cope with these issues facing efficient deployment of LoRa-based industrial systems. Knowing the reference LoRaWAN architecture composed of EDs collecting data and the GW playing the role of relaying node between the EDs and application server. In the proposed scheduling scheme, EDs transmit their data during the data acquisition cycle times, also referred to by frames composed of multiple time slots. A logical slot index was assigned to each time slot in the frame to facilitate the scheduling task and help EDs to respect their data TX time constraint. Distinct SF and frequency channels were assigned to different sets of EDs that are grouped regarding the signal attenuation degree to remove interference between them. In addition, mLBT was used to help avoid TX via a busy channel by performing multiple times the channel activity detection during the same time slot. The superiority of the packet delivery rates of the proposed scheme regarding the existing approaches was analytically and experimentally demonstrated. 

A cognitive model was suggested in \cite{Muthanna2021a} to ensure efficient multiple access in LoRa-based IoT networks, where the interaction between the user traffic parameters and the network operation quality was analytically modeled. Thus, the heterogeneous users' conditions were considered when tuning the network parameters regarding the bit error probability. The authors of \cite{Pham2021} performed extensive real-world experiments to design a reliable and clear channel assessment procedure which is an essential component of CSMA mechanisms. Experiments were performed to understand the concurrent TX mechanism at the MAC layers for a dense LoRaWAN configuration, leveraging the capture effect and channel activity detection properties. 

Cognitive radio was considered in \cite{Mroue2021} to assist LoRa-based IoT applications in which EDs cooperate with each other to determine the primary user status with the help of a fusion center and perform spectrum sensing. Additional energy is consumed during spectrum sensing and reporting to  monitor the primary user status efficiently, which requires the development of a new scheme to manage the data exchange between these EDs. This situation motivated the authors of \cite{Mroue2021} to design a new protocol that defines the nodes eligible to perform spectrum sensing efficiently based on their residual energy. 

\subsection{Slotted MAC}

Multiple solutions were discussed above to decode the collided LoRa signals using channel allocation or carrier sensing. These solutions were proposed as alternatives to the pure Aloha medium access mechanism adopted by LoRaWAN, which suffers in terms of scalability. A frequency or logical channel allocation mechanism was suggested to avoid collision between concurrent TXs. For example, each ED could use a different frequency channel to transmit its data or employ the logical channel defined by a combination of LoRa features such as the SFs, BW, and others. This solution solved the problem of concurrent TX in dense configuration regarding energy efficiency, end-to-end delay, and packet delivery rate. Due to the limited frequency and logical channel possibilities, this solution performs well for a LoRa network capacity under a threshold related to the number of allowed channels. Carrier sensing was also proposed to allow the EDs to sense the medium before transmitting their data, which reduced TX collision. However, additional energy will be consumed to perform sensing, which is not beneficial for the LPWAN network comprising battery-powered EDs. This section discusses another solution to overcome the issues of LoRaWAN scalability, namely the slotted MAC protocols.

The authors in \cite{Beltramelli2021a} proposed an analytical model to evaluate the performance of pure Aloha alternatives, such as slotted Aloha (S-Aloha) and non-persistent CSMA in terms of throughput and energy efficiency while capturing the LoRa PHY layer characteristics. The aim was to emphasize the impact of these random access mechanisms to improve the throughput of LoRaWAN regarding energy consumption and LoRa modulation properties when interference occurs. Thus, the proposed model can be used to design a random access mechanism for different LoRa network scenarios. A method to decode superposed LoRa signals was suggested and integrated into the MAC protocol \cite{8678478}, retrieving the entire frame from desynchronized signals without any loss. Beacons and slotted MAC were used to alleviate the issue of desynchronization and allow multiple EDs to transmit simultaneously using the assigned slots. When the proposed method cannot decode some LoRa signal frames, a CRC decoding scheme was used to cope with that issue.

Although slotted MAC protocols have been broadly investigated in the literature for decades and many real-world applications depend on them, the distinctive LoRa PHY layer characteristics, and the duty-cycle restrictions imposed in sub-GHz ISM bands, render the design of new time-slotted MAC layers suitable for LoRa networks a difficult problematic. The authors of \cite{Zorbas2020a} discussed the specific LoRa characteristics that should be considered when designing time-slotted MAC protocols, and they summarize the current enhancement in this field. A timeslot scheduling scheme that captures the requirements of LoRaWAN-based machine-type communications was presented in \cite{9217770} for massive deployment to increase the number of EDs that can transmit over a single GW. The proposed scheme does not require additional downlink TXs or synchronization between EDs and the GW while considering the particular characteristics of the network such as multi-channel diversity, quasi-orthogonal DRs, and traffic periodicity. A new decoding scheme was incorporated into the slotted MAC protocol \cite{9217247}, which allowed decoding symbols of multiple colliding frames. This decoding scheme compares symbols at each subslot (instead of each symbol end), increasing its ability to decode collided frames. Besides the decoding scheme, the proposed slotted MAC protocol includes retransmission management.

 In \cite{Beltramelli2021b}, LoRa EDs synchronization was proposed to efficiently manage uplink communications resources in slotted MAC protocols using the out-of-band time-dissemination. The use of time-slotted MAC protocols results in the overhead of synchronizing and scheduling the TX, which can be easily afforded regarding the duty-cycle restrictions imposed on the LoRaWAN and the availability of downlink communications. To deal with the overhead, the authors of \cite{Zorbas2020} suggested that EDs autonomously manage their TX and self-organize their slot position in the frame. Moreover, a single slot in each frame is reserved to ensure synchronization and acknowledgment. The aim of \cite{9668447} was to equalize the success probability over the EDs distributed in the network using an S-Aloha mechanism to transmit data regardless of the SF and the source node location. The experimental approach in \cite{Garrido-Hidalgo2021} was based on a central entity to assign slots and ensure synchronization between EDs. Such as, the slot length in the frame was determined using real-world measurements of the clock drift.

The authors of \cite{8588859} demonstrated that S-Aloha could be deployed on top of the traditional LoRaWAN without requiring changes using a new distributed synchronization approach. An adaptive medium access scheme was designed in \cite{Fernandes2021} to ensure the best link quality between EDs and GWs, by adapting LoRa PHY layer parameters and leveraging the coordination ability of beacons regarding duty-cycle restrictions for EDs and GWs. A new MAC protocol was designed in \cite{Triantafyllou2021} to enhance collision avoidance between concurrent TX leveraging fairness by adopting GWs that periodically broadcast beacon frames to synchronize and coordinate between nodes. For reliable networks, the proposed protocol allows EDs to transmit randomly over a specific channel using various CF values based on the scheduling information in the beacon frame. Due to the limited availability of EDs-GWs connectivity, data TX should be scheduled to capture the windows availability of GWs. A time-slotted TX scheduling scheme was designed in \cite{Zorbas2021} to  optimize the data TX time efficiently regarding the issues mentioned above related to the LoRaWAN architecture.

\subsection{Key Insights.}
The limited frequency and logical channels enabled by the LoRa physical parameters should be efficiently managed to prevent the collision of concurrent TX. In this direction, researchers have focused on designing new virtual channel allocation approaches using the game theory while considering the waiting time reduction and ensuring the ED requirement in TX time \cite{Shen2021a, 9149006, 9205428}. Machine learning has also been suggested to automatically manage the TX timing and avoid unnecessary, redundant packet TX caused by widows mismatching between the EDs and GWs \cite{Kaburaki2021}. These solutions assume a perfect SF orthogonality, which is impractical, and the same event should be correlated to reduce the uplink load. Most solutions suggested in the literature rely on throughput to evaluate performance. However, the throughput does not necessarily include only goodput; duplicated packets may be transmitted, resulting in BW waste and useless energy consumption. Therefore, goodput should be considered instead of throughput when evaluating the proposed solutions. 

Moreover, new network topologies have been suggested to increase the LoRaWAN capacity and extend its limited coverage area due to the traditional star topology based on one-hop communication \cite{AsadUllah2019}. In this direction, multi-hop communication where collected data are off-loaded onto multiple subnetworks has been adopted to build scalable LoRaWANs while considering the LoRa PHY-layer standard \cite{8637935}. Although the efforts provided to design new routing protocols \cite{8253754, Valtorta2019, Farooq2020, 9297284} are suitable for LoRaWAN based on battery-powered EDs, new protocols should be proposed considering the fairness between EDs in terms of throughput and energy consumption. In addition, machine learning techniques can be applied to manage the forwarding task to avoid draining the energy of forwarding nodes by splitting their tasks between multiple EDs based on their residual energy.

Furthermore, different carrier sensing schemes have been suggested to efficiently access LoRaWAN communication channels and avoid high concurrent TX collisions due to adopting the Aloha access scheme in the traditional LoRaWAN \cite{8376997, 8966182, 9142799}. The listen-before-talk mechanism showed promising results in reducing concurrent TX collisions. However, this process drains the energy of EDs due to the additional energy spent to sense the channel  \cite{8422800, 8766700}. To cope with this issue, a hybrid MAC protocol based on slot scheduling and mLBT was suggested \cite{Hoang2020} to ameliorate LoRaWAN channel access while reducing the energy consumed during the sensing process. Otherwise, only EDs with high residual energy can be selected to perform channel sensing before the TX \cite{Mroue2021}. 

Nevertheless, efforts are required to adapt the existing slotted MAC protocols into the characteristics of LoRaWANs, such as the specifications of the LoRa PHY layer and duty-cycle restrictions imposed on using free ISM bands \cite{Zorbas2020a, Fernandes2021}. Despite the endeavors provided in this direction \cite{Beltramelli2021a, 9217247, 9668447, Beltramelli2021b}, some drawbacks showed up, such as the need for synchronization, which is not an easy task due to the duty cycle restriction and long time-on-air. This results in the overhead of synchronizing and scheduling the TX as beacons broadcast periodically \cite{Triantafyllou2021}.

\section{Scalable LoRaWAN Use Cases}
\label{sec:Applications}
As stated in the literature, LoRa communication systems have been widely considered to build diverse IoT applications due to their long-range communication, low power consumption, and immunity against interference or concurrent TX collisions. This consideration is due to the CSS modulation and its multiple features, such as the SF, TP, CR, DR, BW, and CF, managed by an ADR mechanism for efficient wireless communication based on battery-powered LoRa devices. However, due to the rapid growth of the network density across many LoRaWAN applications, scalability is becoming a central problem for LoRa-based massive IoT applications to keep it within the scope of LPWAN technology. Previous sections describe different solutions proposed in the literature to enhance LoRaWAN scalability. This section describes some of these applications requiring a scalable LoRaWAN.

\subsection{Space-to-Ground Communications}
\begin{figure}[t]
    \centering
    \includegraphics[width=0.98\linewidth]{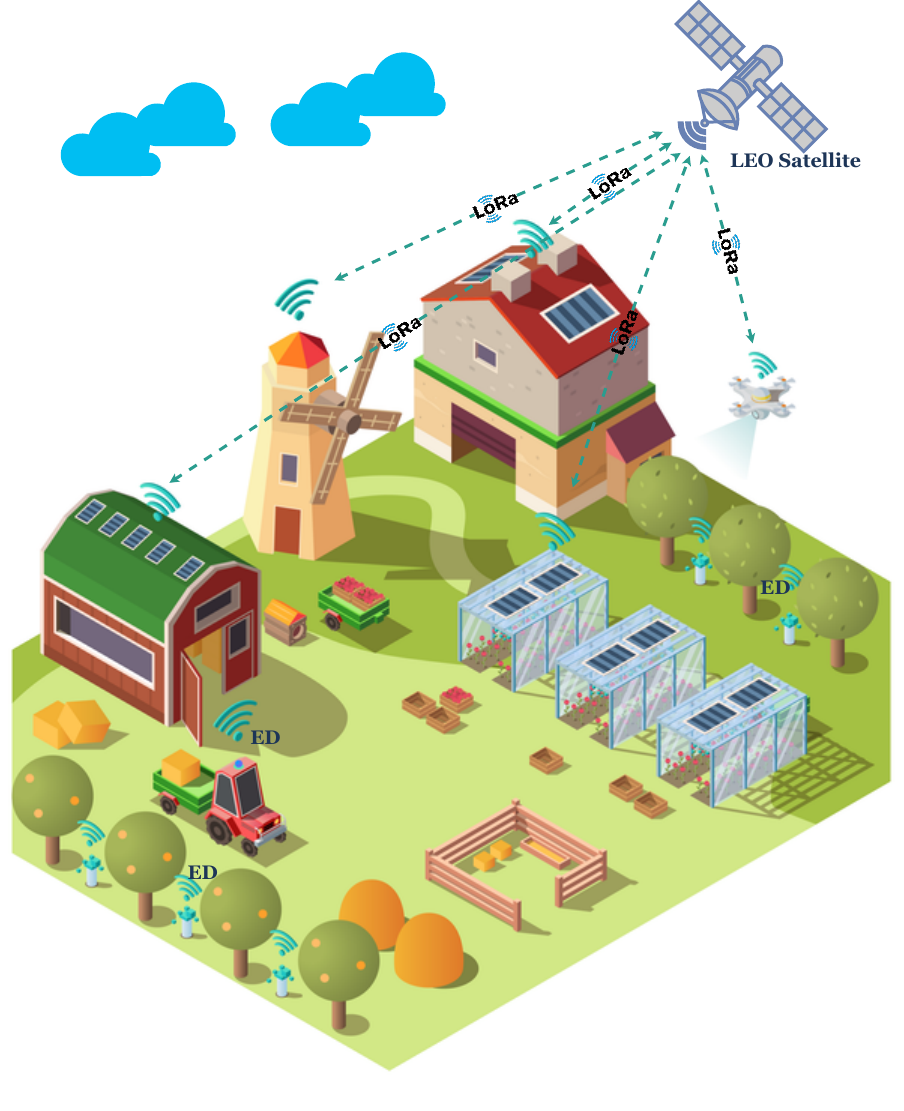}
    \caption{An example of LEO satellite used as GW for LoRa-based smart farming, where EDs deployed in the target area transmit their collected data using LoRa communication into a CubeSat flying in a LEO.}
    \label{fig:space_things}
\end{figure}

The LoRa network has been proposed in the last few years as a leading technology enabling wireless network service in space due to its power in providing long-range communication and the great interest of the research community to establish low earth orbit (LEO) satellites (Fig.~\ref{fig:space_things}). Unlike the enormous traditional satellites, nanosatellites weighing only 1 to 50 kg can be launched into LEO at around 2000~km from the ground. The number of these nanosatellites launched into a LEO has  considerably increased recently due to the advancements in computer and electronic technology. Thus, LEO nanosatellites have attracted substantial attention from the research community as promising alternatives for large-scale coverage in massive IoT networks, providing direct or indirect satellite connectivity \cite{s21217099}.
Instead of developing satellite communication (considering all protocol settings and the RF modulation used to enable robust space-to-ground communication), LoRa has been deployed to build wireless network service in space to allow easy and low-cost management of these nanosatellites. In \cite{techavijit2018internet}, the authors designed a high-quality wireless network in space to connect and manage these nanosatellites from the ground via an internet connection. A study \cite{bashir2021low} analyzed the performance in terms of resistance to Doppler shift packet loss of the newly launched 3U CubeSat, called the Satish Dhawan Satellite, which carries a LoRa transceiver. 

The adaptability of a LoRa  network for LEO satellite IoT deployment was analyzed in \cite{8766462}, according to its characteristics, including LoRaWAN architecture, LoRa node activation mode, the access mechanism for the communication channel, and BW. In addition, some enhancements were proposed to deal with these inapplicable characteristics. The Doppler effect on the LoRa modulation adopted for ground-to-space communication was presented as a major issue facing the advancement of the CubeSat industry. Multiple studies have dealt with this issue \cite{8723123, doroshkin2018laboratory, bashir2021low}. The authors of \cite{doroshkin2018laboratory} identified the Doppler frequency shift resulting from the high speed of these nanosatellites passing over a ground station. This study consists of a laboratory experiment on the LoRa modulation under conditions similar to CubeSat in an LEO regarding ground communication. The aim was to study the immunity of LoRa modulation against the Doppler effect in such conditions. Simulations in \cite{Cao_2021} indicated that the maximum Doppler frequency shift in LoRa communication in LEO CubeSat is about 22~kHz, whereas the maximum Doppler rate reaches 270~Hz/s. Further, a 100\% packet error rate was reached when using an SF of 9, whereas the Doppler rate significantly increased.

In the same way, additional outdoor experiments were conducted \cite{8723123} to study the usefulness of LoRa communication technology in LEO CubeSat radio communication systems. These CubeSat systems face satellite-to-earth radio channel obstructions due to the rapid change in the Doppler effect frequency shift. This degradation appears when the nanosatellite directly above the ground station uses an SF of 12. In addition, the orbit altitude greatly influences the duration of the communication session, as concluded by this study. As explained, GWs are critical components of a LoRaWAN. In this context, the authors of \cite{Zhang2021_2} suggested using ZYNQ SOC to build a space GW for space-based IoT communications to fit the non-operating system requirements of the nanosatellite. This chip serves as software, hardware, and input/output programmability. Different LoRa device configurations were evaluated in \cite{Fernandez2020} to identify the most suitable configuration that meets the space-to-earth channel requirements. Moreover, software-defined radio experiments were conducted to evaluate the performance of the LoRa modulation under harsh conditions concerning the ionospheric scintillation effect. 

A set of IoT devices may be covered under a satellite constellation, which imposes the need to design efficient methods to access the ground-to-space channel. Thus, an intelligent traffic load distribution strategy was designed in \cite{s21217099}. Erasure probabilities employed in this strategy enhance the  overall system throughput by managing the traffic load allocation for different LEO constellation positions regarding IoT clusters. As studied in \cite{Lapapan2021}, access strategies for a LoRa multi-channel should deal with the Doppler effect CubeSat radio communication. In this work, LoRa immunity against the Doppler effect was studied through simulation and laboratory experiments to identify the applicability of the LoRa modulation under such harsh conditions. The sparse satellite constellation allowed in \cite{Fraire2020} used specific LoRa configurations as a maximum clock drift feature of LoRa devices to reduce direct-to-satellite IoT communications. Moreover, when focusing on the communication between IoT devices and GWs on satellites, the security, and privacy of transmitted data should be considered, especially when we know that most use cases of this system rely on critical infrastructure. A security scheme was suggested in \cite{Anantachaisilp2020} based on the Advanced Encryption Standard (AES) to secure the transmitted data in the private LoRaWAN-server via satellite GWs. 

The LoRa receiver designed in \cite{Ivansyah2021} for IoT devices in remote areas, such as islands or mountains, transmits and receives data from CubeSat. The receiver has an SX1276 chip to process the received passband signals. A low $R_b$ is allowed using SFs of 7 and 12, respectively, equal to 5468.750 and 292.969~bps. Due to the insufficient space allowed by the CubeSat geometry to embed a LoRa antenna, for the dependency between wavelengths and antenna size, the authors of  \cite{Zalfani2021} proposed planar antennae, where the meander line and the ground plane are placed on the same side. The authors of \cite{Suraj2019} suggested benefiting from the rocket debris of various polar satellite launch vehicles (PSLVs) revolving in different space orbits to build a low-cost and flexible software-defined networking platform. Instead of destroying the PSLV rocket and throwing it into the ocean, PS4 PSLV Stage 4 retains this debris in its initial orbits, weighing around a ton. This work aims to enable inter-PS4 and ground-to-PS4 communication through LoRa technology. Performed simulations showed the feasibility of the proposed network to ensure high connectivity besides the enhanced throughput. LoRa-based CubeSate design and challenges were the focus of \cite{Perez2020, Prokopyev2021, Akyildiz2019}, such as radiomeTry and vegetation Analysis (RITA) payload with a customizer LoRa transceiver integrated into a 3U satellite \cite{Perez2020}. In \cite{Prokopyev2021}, the authors analyzed the NORBY CubeSat nanosatellite launched into an LEO on September~28,~2020. The internet of space things based on CubeSats to enable low-cost global connectivity was introduced in detail for the first time in \cite{Akyildiz2019}.

\subsection{LoRa-based Localization Systems}
Positioning in the massive IoT is becoming relevant for current and future IoT applications based on LPWAN technology, often providing a substantial added value. Due to the efforts to enhance its scalability and provide connectivity for massive IoT networks, the LoRaWAN is becoming the leading technology to build IoT positioning systems for smart cities. This section discusses the latest work on LoRa-based localization systems published from 2019 to the present, summarized in Table~\ref{tab:localizations}.

An experimental study was conducted in \cite{Khan2021} to evaluate the efficiency of wireless communication techniques to perform indoor localization. This study covers Bluetooth low energy (BLE), Wi-Fi, and LoRa, considered the most suitable communication techniques for indoor locations due to their low energy consumption. However, an efficient localization system should be robust and cost-effective and provide a real-time and accurate localization decision. Based on the outcome of this study which used the RSSI to calculate the node coordinates, Wi-Fi is the best technique to provide an accurate localization, whereas BLE is the worst. Between them, LoRa communication systems localize devices with an average error of 0.62~m. The RSSI-based localization system faces many problems, such as signal fading related to the land-cover types or labor-intensive problems, in providing accurate localization in the outdoor environment. To address this, the authors of \cite{Lin2020} used high-resolution satellite images to manage the path loss by identifying different land-cover types using the collected images of the target area. By profiling the environmental parameters, SateLoc allows the cooperation of multiple GWs to provide a highly accurate outdoor localization with an error of 47.1~m. All the localization schemes in this section are summarized and classified in Table \ref{tab:localizations}.

Localization algorithms can be categorized into two categories as explained in \cite{Janssen2020}. The first comprises range-based algorithms, where the RSSI value received from the GW can be converted into a distance using a specific path-loss model. The combination of ED distances from GWs help  calculate the localization. Second, fingerprint-based algorithms in this location category  of EDs are approximated by comparing the current RSSI values of the device and the previously collected RSSI values of known locations. This approximation can be provided using machine learning models trained on labeled RSSI datasets as indicated in \cite{Janssen2020}. Several methods can be deployed in a range-based localization algorithm to provide an accurate devices location as proposed in \cite{Aernouts2020, 10.1145/3517245} where the time difference of arrival (TDoA) and angle of arrival (AoA) are combined for time-based localization. The first relies on the propagation time to determine the device coordinates, whereas the other uses the RSSI values received from multiple GWs and calculates the corresponding AoA to provide a location estimation for the ED. This paper demonstrates that combining these two methods requires using only two GWs to achieve an efficient location with a mean estimation error of 339m for line-of-sight (LoS) signals and 159m for a non-line-of-sight (NLoS) signals.
\begin{table*}[t]
  \centering
  \renewcommand{\arraystretch}{1.8}
  \caption{Performance comparison of the recent LoRa-based localization systems}
  \begin{tabular}{l c c p{6em} p{8.5em} p{6em} p{10em} p{12em} }
  \toprule
  \textbf{Paper} &\textbf{Year} & \textbf{Environment} & \textbf{Technique} & \textbf{Performance} & \textbf{Evaluation} & \textbf{Advantages} & \textbf{Disadvantages}\\
  \bottomrule
   \cite{Khan2021} & 2021 & Indoor & RSSI range-based & Wi-Fi, error = 0.54m 
BLE, error = 0.82m
LoRa, error = 0.62m & Experiment & Provides a comparison of LoRa, Wi-Fi and BLE-based localization systems & Unfair comparison since the use case for each should be considered\\
  \hline
  \cite{Lin2020} & 2020 & Outdoor & Satellite images, fingerprint & Error = 47.1m & Dataset \cite{DeepGlobe2018} & High potential for performing large-scale tracking & Needs more real-world evaluations\\
  \hline
  \cite{Janssen2020} & 2020 & Outdoor & RSSI, range based and fingerprint & Range-based error = 700m Fingerprint-based error = 340m & Dataset \cite{data3020013} & Efficient GWs selection strategy & Localization in a predefined area \\
  \hline
  \cite{Aernouts2020}  & 2020 & Outdoor &  TDoA, AoA range-based & LoS error 339m, NLoS error 159m & Simulation \cite{aernouts2019simulating} & Combining TDoA with AoA using only two GWs & Should include more synchronized GWs\\
  \hline
  \cite{Liu2021a} & 2021 & Indoor & Round-trip time, fingerprint & LoS error = 1m & Experiment & Use of particle filter to design a LoS and CDF to evaluate its performance  & Should focus on pseudo-range correction for NLoS\\
  \hline
  \cite{Bnilam2021} & 2021 & Outdoor & AoA range-based & LoS error = 2°, NLoS error = 10° & Simulation & Provide an accurate estimation of the  AoA in an intensive multipath environment for LoS and NLoS & Study restricted to LoRa systems\\
  \hline
  \cite{Vazquez-Rodas2020} & 2020 & Outdoor & RSSI range-based &  LoS error = 7\% of max distance & Simulation & Use of LoRa Pycom RF capabilities & Study restricted to LoPy and FiPy devices\\
  \hline
  \cite{Bouras2020} & 2020 & Outdoor & RSSI range- based & Error about 40-60m & Simulations & Operate in indoor environments (playgrounds, or  shopping malls) & Machine learning can help on improve system performance \\
  \hline
  \cite{SCIULLO2020101993} & 2020 & Outdoor & RSS, PL range-based & Error $\leq$ 90m & OMNeT++ & Enhance emergency request deliverance using multi-hop communication & Phone dependency, simulation displays optimistic results, requires experiments\\
  \hline
  \cite{9169667} & 2020 & Outdoor & PL, range-based & Error $\leq$ 60m & Measurements assessed by simulation & The PL model is designed for a challenging environment & A low number of devices can be simultaneously monitored\\
  \hline
  \cite{KHAN2022102732} & 2021 & Indoor & RSSI range-based &  Wi-Fi, max error = 1.12m
LoRa, max error = 4.63m & Experiment & Suggest PL model for the indoor environment & The study focused on a particular indoor environment\\
  \hline
  \cite{9328764} & 2021 & Indoor & LoRa Backscatters & Accuracy = 89.7\% & Simulation & Provide 3D localization & Real data can improve \\
  \hline
  \cite{8827665} & 2019 & Indoor, outdoor & RSSI & Error = 70.41~m & Simulation, real experiments & Proposed model cover indoor and outdoor & Adopt basic localization algorithm\\
  \hline
\end{tabular}
  \label{tab:localizations}
\end{table*}

In contrast, filtering algorithms such as Bayesian and Kalman filters can be deployed to build an efficient, stable, and accurate localization system. These filters can aid the localization system in providing a more accurate location by solving LoRa indoor localization problems concerning the range error caused by multipath. As discussed in \cite{Liu2021a}, a new particle filter was designed to cope with the LoRa pseudo-range fitting. The authors of this paper provided a fingerprint of the ED using a real LoRa round-trip time measurement and adopted the cumulative distribution function criteria to calculate the quality of the estimated location by comparing it with the ground truth. The AoA of the received signals was estimated in a real-life urban environment by developing a complete system comprising of hardware and software solutions \cite{Bnilam2021}. This work aims to estimate the AoA from the received signals, which are highly correlated using a space-alternating generalized expectation-maximization algorithm. This system estimates the AoA of the received signal for  LoS and NLoS with an average error  of 2° and 10°, respectively. The LoPy and FiPy are two low-cost LoRa devices used in \cite{Vazquez-Rodas2020} to build a low-cost local positioning system based on the RSSI to fit some IoT application requirements that are not allowed by the GPS. These applications require the deployment of multiple small resource-constrained devices. The proposed system benefits from Pycome RF capabilities, which were evaluated in an outdoor rural environment and showed an average error of the estimated position of around 7\% of the maximum distance between EDs.

The LoRa-based localization systems can be designed for critical applications such as search and rescue (SAR) operations, which require fast localization of those involved in accidents in a harsh environment \cite{etiabi2022spreading, etiabi2022federated, etiabi2023federated}. Some standards exist for SAR missions in snowy environments, such as ARVA and RECCO, though these standards suffer from their limited radio range. Their limited performance motivated the research community to seek an alternative offering a long communication range with an extended lifetime. For this purpose, multiple research works have been published in the last few years. The authors of \cite{Bouras2020} proposed analyzing the behavior of the LoRaWAN channel while using trilateration methods and RSSI values to provide the localization of the target person in the concerned area with an error average from 40– to 60~m. 

The authors of \cite{SCIULLO2020101993} focused on the mobile emergency management system that allows people to send emergency requests from their cell phones to rescue teams. Despite, the pervasiveness of these devices, LoRa-based localization is required to support the effectiveness of mobile-based rescue systems in critical scenarios. Emergency requests sent from a cell phone can be forwarded by multiple peers equipped with LoRa transceivers using BLE until reaching the rescue teams. This process increases the probability that a request message is successfully received at its destination, taking advantage of LoRa interference immunity and the long communication range. This model implements the LoRa-based trilateration technique and uses path-loss measurements to provide an accurate GPS-free localization with an average error of less than 90m. A radio path-loss model has been developed in \cite{9169667} based on real path loss measurements in a harsh mountain environment to estimate the coordinates of a lost hiker either at the bottom of a canyon or on the top of a snowy mountain. This study revealed that localization systems based on LoRa path-loss estimation outperform those existing in this field, namely ARVA, by achieving a five times longer radio range.

The exact location of objects is not always required, especially in some indoor localization systems. In this application, we only need to know in which place, room, or hall the object is present, which tolerates a considerable estimated error range. In this direction, the author of \cite{KHAN2022102732} suggested a two-phase cell localization algorithm based on the LoRa trilateration algorithm, consisting of finding the cell location of the target object. The first phase begins with analyzing the path-loss exponent for all anchor nodes using the received RSSI value, whereas the second phase involves dividing the target space into virtual cells to determine the cell where the object is present. This study considers two wireless channels: Wi-Fi and LoRa, where the proposed method had a mean error of 1.12m and 4.63m. A LoRa Backscatters-based room-level localization system was designed in \cite{9328764}. The system uses LoRa transceivers to transmit an RF signal by modulating the received signal. This system comprises multiple LoRa receivers deployed in different rooms, a LoRa transmitter at a central point (reference point), and the backscatter device carried by the object. The RSS values are compared between all the receivers to determine the position of the LoRa backscatter devices, whereas machine learning enhances the system performance. Based on the linear discriminant analysis, this system achieved a localization accuracy of 89.7\% in a real-life scenario. Multiple algorithms of localization based on the LoRa RSSI values are proposed in \cite{8827665}, to face the issue of Gaussian and non-Gaussian noise by reducing their influence on the localization accuracy. Through simulation and real experiments, this model demonstrated its ability to perform well in indoor or outdoor environments  \cite{10.1145/3447993.3483256}. 

\subsection{Smart Building}
Smart building is the newest application field of LoRa communication systems to enable efficient IoT connectivity inside crowded buildings, known for its high concurrent TX collision resulting from the high connectivity demand. Due to scalability, low cost, and easy deployment, LoRa technology is becoming the leading technology in developing smart buildings. A comparison study in \cite{Fraile2020, 10.1145/3274783.3275218} demonstrated the high capabilities of LoRa technology to enable school monitoring compared to IEEE 802.15.4 wireless technology. Besides the low-cost IoT deployment, LoRa was implemented on Arduino-based hardware to build robust and reliable monitoring systems due to its low energy and random MAC protocol. The authors of this work provided a rich dataset produced from six school buildings using 49 devices.

The LoRa communication technology has already been proposed as a part of a smart healthcare system \cite{Trigo2020}. It enables patient tracking inside a hospital comprising multiple buildings connected via underground tunnels. Intra-hospital transport of patients, especially those admitted in intensive care units, is one of the major challenges facing the efficient deployment of the internet of medical things. In this case, patients require periodic transportation between units within the hospital buildings for diagnostic reasons. Consequently, determining their location is crucial for efficient transportation and coordination between units, which is evermore challenging for hospitals with massive infrastructures. 

Advanced meter infrastructures make buildings more intelligent helping design smart buildings throughout and allowing efficient and intelligent electricity power consumption. The authors of \cite{Gallardo2021} designed a LoRa-based architecture to manage the electricity power intelligently inside a residential grid and evaluated their system regarding different factors such as energy efficiency, packet delivery ratio, and throughput. This architecture is based on the ADR features of the LoRaWAN to address the dynamic system behavior and benefit from a scalable LoRa network. The study aimed to determine the right place to put GWs for cost-effective and efficient deployment. The LoRaWAN is used in \cite{Klaina2020a} as a low-cost infrastructure for electric vehicles for grid communications in smart cities. The LoRaWAN is proposed for this application due to its long communication range and its ability to reduce power consumption  by allowing efficient coordination between electric vehicles and the corresponding aggregators or enabling low-cost communications between nodes. 

The LoRa network has been suggested for developing highly critical applications, such as for indoor air quality \cite{Pereira2020}. Specifically, this network is integrated into radon risk management in public buildings with high density, as radon is a radioactive natural gas classified by the World Health Organization as the most dangerous gas, which can cause lung cancer. Radon gas naturally occurs and accumulates in the indoor environment, such as in a school. Thus makes LoRaWAN plays a crucial role in reducing public health risks associated with this pollutant. Another critical use case of the LoRa communication system is monitoring elderly people and following up on their health conditions, especially those isolated for the coronavirus disease 2019 (COVID-19) infection. The proposed system in \cite{Lousado2020} integrates low-cost devices connected to the IoT network via LoRa communication for data processing and communication. Various sensors are incorporated into this system to collect critical data that allow authorized entities to obtain real-time information on health and the urgent needs of the system users. In addition, sensor networks can be employed to monitor older people using LoRa GWs for cost efficiency. 

\subsection{Environmental Monitoring}
Often, LoRa technology is used for monitoring harsh environmental conditions as part of smart city systems, taking advantage of its large coverage and low TX rates (low energy consumption). Furthermore, the technology adapts its SFs and DRs to fit the communication environment conditions, dynamically managing link quality. The authors of \cite{Ferreira2020} used the RSSI, SNR, and packet delivery ratio to study the propagation of  LoRa signals in urban, suburban, and forest environments for a mobile node scenario. Monitoring wildlife and integrating it into smart agriculture are recent use cases of the LoRa communication system \cite{Ojo2021}. The design for a wildlife monitoring system was provided in  \cite{Ojo2021} for IoT animal-repelling devices. The aim was to overcome the issues while incorporating cellular and short-range wireless technologies for this application type. In this study, the RSSI values, SNR, and packet delivery ratio are used to characterize the communication link to evaluate the performance of the proposed solution.

 The environmental radiation monitoring system designed in \cite{GallegoManzano2021} comprises numerous LoRa EDs deployed in CERN sites that periodically measure  radiation levels and send the data to a LoRaWAN server via LoRa GWs. This approach automates controlling the radiation level in conventional waste containers and provides a real-time warning of potential radiation release. Likewise, the authors in \cite{Fort2021} proposed to integrating the LoRaWAN into a carbon monoxide measurement system in smart cities to monitor the carbon monoxide emission with a cost-effective and increased lifetime of the system. Safety is the aim of the LoRa-based use case in \cite{Jiang2021a}, where LoRa devices were deployed alongside high-voltage TX lines for real-time temperature monitoring. LoRa technology was used in this scenario as an alternative to the legacy monitoring systems, such as manual or drone-based inspection and short-range wireless technologies for its long communication range. Moreover, multi-hop communication was adopted to allow broader coverage for data TX between adjacent nodes.

The LoRaWAN network was suggested in \cite{Kombo2021} to manage groundwater resources by providing real-time information on groundwater levels. Supported by its low cost and low power consumption, the LoRa-based groundwater management system was revealed as the most suitable technology to deploy in developing regions. The usability of the LoRa technology to monitor underground activities was assessed in \cite{DiRenzone2021}. The study focused on discussing the ability of the LoRa PHY layer and LoRaWAN to communicate data from underground sources to the aboveground base station. The analyses were performed considering three soils (gravel, sand, and clay) at 50~cm depth, acquiring SNR and RSSI values to evaluate the TX performance. In addition, LoRa through-wall sensing was explored in \cite{Zhang2020} to discover new sensing opportunities using LoRa transceivers. Recently, in the military context, LoRa build IoT applications for decision-making by collecting data from the battlefield \cite{DBLP:journals/corr/abs-2108-02281}. The role of LoRa communication in this application provides resilience and survivability for the military network, benefiting from preinstalled IoT devices and their capacities.

\section{Discussion \& Future Research Directions}\label{sec:Future}
Existing research has demonstrated that massive IoT requirements in terms of connectivity exceed the legacy of LoRaWAN capabilities due to its limited features, such as the SFs (7-12), BWs (125; 250; and 500 kHz), and CR ( 4/5 4/6 4/7 4/8). The standard ADR to manage the massive connectivity demands and low impact of the CSS modulation to ensure efficient wireless communication in ultra-dense deployment is challenging. This problem resulted in LoRa signals interference and collision between concurrent TX, which drastically reduced the network performance and rendered the deployment of the LoRaWAN in this scenario unsuitable in its traditional version. This problem motivated the research community to propose various enhancements to the LoRaWAN protocol to ensure a reliable and scalable LoRa network. Nevertheless, various open research challenges must still be investigated further.

\subsection{Joint Scheduling and SF Assignment}The SF parameter allows EDs to adapt their TP, TX range regarding the used frequency, and BW. The SFs may be assigned randomly, like the legacy LoRaWAN or could be based on the distance of the EDs from the GW, such as the EDs nearest to the GWs using the lowest SF to transmit their data. In contrast, the faraway ED use the highest SF value 12 (Fig.~\ref{fig:SFassignement}). Other methods assign a single SF to a set of EDs based on their communication link quality or requirements regarding the collected data type. All these methods showed encouraging results in reducing LoRa signals interference and enhancing the network scalability. However, the limited SF combination restricts the enhancement of the LoRaWAN capacity to efficiently connect a high number of EDs in massive IoT deployment. Therefore, the SF assignment may join the TX scheduling in a cross-layer protocol that involves linking the LoRa PHY layer characteristics and LoRaWAN MAC layer capabilities.
\subsection{Efficient Interference Cancellation Schemes} Interference cancellation is implemented on EDs or GWs to decode multiple superposed LoRa signals transmitted using different SFs. Various signal processing methods, such as FFT and discrete Fourier transform, can distinguish between interfered signals. The capture effect may be considered where only the strongest signal is decoded. Furthermore, frame preambles are identified from the received signal to decode them effectively. Additional information should be exchanged, including the SF and CR used to transmit the data, which requires extra energy consumption. These time-consuming methods are unsuitable when dealing with time-sensitive information and increase the cost of receivers not allowed in LPWANs. Therefore, a trade-off between the cost and network performance regarding the received packet rate should be considered when implementing this solution to cope with interfered LoRa signals.
\subsection{Novel Densification Techniques} Cellular networks may inspire researchers to design new densification strategies for LoRa networks concerning the restrictions, such as the duty cycle and low throughput. This solution is traditional concerning the increased demand for connectivity in a crowded urban environment, such as downtown. Joining the GWs densification with the previous solution mentioned in this work can significantly increase the capacity of the LoRaWAN to connect more EDs. However, some architectural changes may be required to adapt the legacy LoRaWAN to its new enhancement, requiring specific attention when designing a densification strategy to facilitate the coordination within and between cells and avoid additional collision between concurrent GWs TX.
\subsection{Optimized Windowing \& Channel Allocation} Resource management schemes are primarily employed to fit the TX windows of EDs to the RX windows of the GWs. The RX windows of the GWs are not always available since TX should be opened to acknowledge the received data. Additionally, due to energy efficiency requirements, EDs can not keep their TX and RX windows always open. Therefore, this mismatch should be solved considering the network density, which restricts the impact of these methods to enhance the LoRaWAN capacity. Thus, the logical and frequency channels were managed in such a way as to allow simultaneous concurrent TX of multiple EDs. Promising results were observed in this context using smart resource assignment schemes. Furthermore, mathematical tools such as game theory can manage resources by modeling the interaction between the EDs and GWs and leveraging the various game types depending on the target scenarios. Alternatively, reinforcement learning can also be deployed as an optimization tool.
\subsection{Relaying and Routing} The conventional LoRaWAN adopts the star topology to connect EDs to the network through peer-to-peer connections with the GWs. The EDs are deployed randomly or following a specific strategy using the TP and SF to transmit their collected data via a frequency channel. Massive direct connection to the GWs results in collisions between concurrent TXs and drastically decreases network performance regarding the received packet rate or communication latency due to multiple retransmissions of the collided data. Hence, multihop communication can be adapted to deal with these issues where EDs far from the GW can rely on intermediate EDs to transmit their data. Otherwise, the new topology may be designed to seek fairness between EDs in terms of link quality, SFs, or energy efficiency to increase the network lifetime. Consequently, new routing protocols should be designed to assist this enhancement which can be performed conjointly with the GW densification for efficient network topology design. Furthermore, a set of EDs with low residual energy can rely on a cluster head with high residual energy and ensure the forwarding task between the cluster members and GW.  
\subsection{Energy-Efficient Carrier Sensing} The listen-before-talk mechanism is leveraged to enhance the LoRaWAN channel utilization and efficiency. The EDs can sense the medium and decide whether the channel is idle or occupied by another TX using the same SF and signal characteristics. Carrier sensing is an energy-consuming process that raises the interest in carefully using this method, especially when dealing with the battery power of EDs. However, due to the long time-on-air of long-range communication that decreases the performance of carrier sensing methods, multiple channel sensing may be required to benefit from its advantage efficiently. Moreover, data TX via multi-hop communication should be enabled in the subsequent medium access protocols. Adapted CSMA/CA protocols were suggested for this, considering a portion of hidden nodes with promising results in terms of collision avoidance regarding various sensing ranges. Despite provided efforts, energy consumption remains the main issue facing the efficient deployment of CSMA methods in LoRaWAN. 

\subsection{Machine Learning for MAC} 

Slotted MAC protocols are suggested as an alternative to pure Aloha-like protocols adapted by LoRaWAN due to their limited scalability. Nevertheless, designing an efficient time-slotted MAC protocol faces severe challenges due to the specific characteristics of LoRa and the duty cycle restrictions imposed in sub-GHz ISM bands. Further, decoding schemes, such as the CRC may be incorporated to help decode symbols of multiple colliding frames. Thus, an overhead of synchronizing and scheduling the TX was raised due to the duty-cycle restrictions imposed on the LoRaWAN and the sparse availability of downlink communications. Autonomous and distributed scheduling schemes were considered to overcome these problems, such that EDs collaborate to self-organize their TXs and avoid collisions regarding the requirements and communication conditions for each node. Deep reinforcement learning opens a new research perspective to solve this issue. Thus a DRL agent can be designed and trained off-line to manage MAC layer resources and efficiently run onboard the EDs. The aim is to allow local decision-making on MAC layer resources to enable efficient communication while considering the behavior of other EDs and the limitation imposed by the LoRaWAN specification. The MDP environment should be designed efficiently to incorporate all characteristics of the considered system model to obtain the full benefit from this technique. Furthermore, the action and state spaces require meticulous interest due to their impact on the reward convergence of the agent during the training process.

\section{Conclusion}

Although LoRa is a promising technology for massive IoT networks to provide reliable long-range communication, it suffers from signal interference and concurrent TX collisions. Therefore, this survey examined the scalability challenges facing the deployment of the LoRaWAN in ultra-dense IoT networks where massive LoRa connectivity is required. Moreover, we reviewed the state-of-the-art solutions in the PHY and MAC layers to address the scalability challenges. We discussed the literature on LoRaWAN deployment for massive IoT connectivity requirements, where modulation, channel characteristics, interference, and collision are the major issues. Most existing solutions addressing these challenges leverage LoRa features to extend capacity, whereas others designed new relay/routing protocols or relied on signal processing techniques to cancel the interference. Moreover, CSMA and S-MAC were employed at the MAC layer to cope with concurrent TX collisions. However, more efforts are needed to develop reliable and scalable LoRa-based massive IoT systems. Therefore, this survey also presented some enticing research directions to assist researchers in designing efficient and scalable LoRaWAN systems.

\section*{Acknowledgement}
This work was sponsored by the Junior Faculty Development program under the UM6P – EPFL Excellence in Africa Initiative.

\bibliographystyle{ieeetr}
\bibliography{references}

\end{document}